\documentstyle[psfig]{mn}
\title[PL and PLC relations for Miras at maximum light]
{Period-luminosity and period-luminosity-colour relations for
Mira variables at maximum light}

\author[S. M. Kanbur, M. A. Hendry and D. Clarke]
{S. M. Kanbur$^{1}$, M. A. Hendry$^{1,2}$ and D. Clarke$^{1}$\\
 $^1$Department of Physics and Astronomy, University of Glasgow,
Glasgow G12 8QQ \\
 $^2$Astronomy Centre, University of Sussex, Falmer, Brighton, East Sussex 
BN1 9QH \\}

\date{Accepted ------. Received ------; in original form \today}

\begin{document}
\maketitle

\begin{abstract}
In this paper we discuss the evidence for a period-luminosity (PL) relation
and period-luminosity-colour (PLC) at
maximum light for Mira variables. We confirm the existence of such relations
in the J, H and K bands, and also based on bolometric magnitudes, for
oxygen-rich (O) and carbon-rich (C) Miras in the LMC. We demonstrate that in
the J and H bands the maximum light PL relations have a significantly smaller
dispersion than their counterparts at mean light, while the K band and
bolometric PL relations have a dispersion comparable to that at mean light.
In the J, H and K bands the fitted PL relations for the O Miras are found to have smaller dispersion than those for the C Miras, at both mean and
maximum light, while the converse is true for the relations based on bolometric
magnitudes. The reduction in dispersion between mean and maximum light is
generally found to be more significant in the J, H and bolometric bands than in
the K band -- in which the mean PL relations display the smallest dispersion.
The inclusion of a
non-zero log period term in the regression fits is found to be highly
significant in all cases except that of the C Miras in the J band, for which
the data are found to be consistent with having constant absolute magnitude.
Moreover, if we consider only C Miras with periods in excess of 250 days,
then the data are also substantially more consistent with constant
absolute magnitude at mean and maximum light in the H, K and bolometric  
wavebands. This suggests the possibility of employing C Miras as standard
candles.
 
We suggest both a theoretical justification for the existence of Mira PL
relations at maximum light and a possible explanation of why these relations should
have a smaller dispersion than at mean light. The existence of such maximum
light relations offers the possibility of extending the range and improving  
the accuracy of the Mira distance scale to Galactic globular clusters and
to other galaxies.
\end{abstract}


\section{Introduction}

Miras are long period variable stars lying on the asymptotic giant branch
(AGB) of the HR diagram with periods
in the range 100 to 700 days. Their masses lie between 0.5$M_{\odot}$
and 3$M_{\odot}$, and their K band mean absolute magnitudes lie in the range
$-5 < M_K < -7$ (c.f. Wood 1995).
Whitelock (1995) reviews a number of reasons for the astrophysical
importance of Mira variables, highlighting in particular their suitability
as distance indicators -- a fact which makes them useful tracers
of galactic structure. The use of Miras as distance estimators relies upon the
existence of period-luminosity (PL) and period-luminosity-colour (PLC)
relations at mean light, which
may be calibrated with nearby stars whose distance is otherwise known and
then applied to more remote objects to estimate their distance.
In e.g. Feast et al. (1989, hereafter F89) PL and PLC
relations were derived for a
calibrating sample of about 50 oxygen-rich (O) and carbon-rich (C) Miras
in the Large
Magellanic Cloud (LMC), using time-averaged mean J, H, K and bolometric 
magnitudes. F89 found that the O Miras displayed a well-defined relations
in the K band, and also based on J, H and bolometric magnitudes, but with a
larger dispersion in these latter three cases. For the C Miras F89
confirmed the existence of a PL relation in the K band.
These derived relations were then applied to determine distance 
moduli to a number of galactic globular clusters.

In this paper, using the same calibrating sample of LMC Miras as in F89
we investigate the evidence for PL and PLC relations at maximum light.
The primary motivation for this work comes from Kanbur \& Hendry (1996,
hereafter KH), who derived V band PL and PLC
relations at maximum light for a sample of Cepheids in the LMC, previously
published in Martin, Warren \& Feast (1979).
KH outlined specific physical reasons why the use of Cepheid maximum light
relations might be preferable to those at mean light, developing the
earlier theoretical work of Simon, Kanbur \& Mihalas (1993).
In a similar manner, in this paper we derive maximum light PL and PLC
relations
for Mira variables and discuss a possible physical justification for their
existence. In particular we consider a physical explanation for the
smaller observed dispersion of maximum light relations when compared with the 
corresponding relations at mean light. For completeness we also consider
PL and PLC relations at minimum light and compare them with their
counterparts and mean and maximum light.

A number of authors (c.f. Sandage 1958, Madore \& Freedman 1991, KH)
have discussed the theoretical justification for the
existence of a PL and PLC relation for Cepheid variables, deriving the
so-called {\em pulsation equation\/},
\begin{equation}
\log P + {1\over 2}\log {\cal{M}} - {3\over 4}\log L + 3 \log T_e =
\log {\cal{Q}}
\label{eq:eq1}
\end{equation}
where $P$, ${\cal{M}}$, $L$, and $T_e$ are the period, total mass,
equilibrium
luminosity and effective temperature respectively of the star and
${\cal{Q}}$ is a slowly varying function of stellar parameters.
Cepheids occupy an instability strip of
finite width in the HR diagram. If a similar situation holds for
Miras (Feast 1989, Wood 1990, Shibahashi 1993), then
this equation can be used to explain the existence of PL and PLC
relations for both Miras and Cepheids since it assumes only the
period--mean density theorem and the Stefan--Boltzmann law. Mira and
Cepheid PL relations arise from the collapse of equation (1) over the
variables $\log {\cal{M}}$ and $\log T_e$. In the case of Miras, however, the 
equilibrium luminosity is a strong function of the core mass
(Shibahashi 1993).
Assuming that the equilibrium luminosity is close to the mean luminosity
over a pulsational cycle, both the range of core masses and total
masses therefore contribute to the scatter in a PL relation at mean
light for a Mira of a given period. We discuss the effect of metallicity
on the scatter of the PL relation in section 4 below.

If it is the case that
\begin{equation}
R_{\rm{eq}} \approx R_{\rm{max}}
\label{eq:eq2}
\end{equation}
that is, the equilibrium photospheric radius of the star is approximately 
equal to the photospheric radius at
{\em maximum\/} light, then, following essentially the same reasoning as in
KH, we can use the period--mean density theorem and the Stefan--Boltzmann law
to write
\begin{equation}
\log P + {1\over2} \log {\cal{M}} -
{3\over 4}\log L_{\rm{max}} + 3 \log T_{\rm{max}} = \log {\cal {Q}}
\label{eq:eq3}
\end{equation}
where $L_{\rm{max}}$ and $T_{\rm{max}}$ denote the luminosity and temperature 
at maximum light. In the case of Cepheids, Cox (1974) provides good evidence 
that equation (2) is a reasonable assumption. Support
for the validity of equation (2) in the case of Mira variables is given in
Wood (1995) and references therein. Assuming equation (2) to be valid for
Miras, equation (3) can then be used to justify theoretically the existence of
PL and PLC relations at {\em maximum\/} light for these stars, as a result
of collapsing equation (3) over the variables $\log {\cal{M}}$ and
$\log T_{\rm{max}}$.

In the Cepheid case, Simon, Kanbur \& Mihalas (1993) showed that at maximum
light the range of photospheric temperatures is significantly smaller than
the range of effective temperatures at mean light. Motivated by this work
KH suggested
that Cepheid PL and PLC relations at maximum light could have significantly
smaller dispersion than at mean light -- a result which was investigated in
detail in KH. A similar effect may not be present for Mira variables,
but another advantage of the use of maximum light is nonetheless
apparent for Miras.
Their pulsations, like Cepheids, are envelope phenomena -- energy 
modulation and amplitude
limitation occurring in the outer envelope.
The maximum luminosity depends on the envelope mass as well
as the core mass. In equation (2), on the other hand, the
equilibrium luminosity is strongly
dependent on the core mass (Shibahashi 1993). Thus the quantities
$P$, and ${\cal{M}}$ in equation (1) have
dependencies on both the envelope and core mass, whereas $L$ is
dependent strongly on the core mass. All the quantities in equation (3) have
dependencies on the core and enevlope mass. We conjecture that,
even if the range of $T_{\rm max}$ were no different to the range of
$T_{\rm eff}$, this situation could lead to Mira PL and
PLC relations at maximum light to have smaller dispersion than
rheir counterparts at mean light. Further work is needed to examine this
proposition.

This paper is organised as follows. Section 2 describes the LMC calibrating
data, and the methods used to derive PL and PLC relations and evaluate their
statistical significance. In sections 3 and 4 we present our results
for PL and PLC relations respectively, which
are then discussed further and compared with those of F89 in section 5, 
highlighting some important consequences for the use of Miras as probes of 
galactic structure. Finally, in section 6 we present our conclusions and 
possibilities for further study.

\section{Data}

The data used in this study were taken from Glass et al (1990), which
was also the primary reference for the analysis of F89. These data consisted
of multi-epoch observations at a number of wavelengths of a large sample of
O and C Miras in the LMC. O Miras are oxygen-rich objects whilst C Miras
are carbon-rich; the classification of each star as a C or O Mira can be
made from spectral type (if known) or from colour measurements and we
adopt the same classifications as those published in F89. In addition we
adopt the Mira periods as given in Glass et al (1990); since the
average number of epochs of observation for each star in the Glass et al. 
study was more than eleven, with good phase coverage, it is unlikely that the 
published periods are subject to any significant uncertainty.

The data for the O and C Miras are summarised in Tables 1 and 2 respectively. 
These Tables list the star name, its period, taken from Glass et al (1990), 
and the mean, maximum and minimum
magnitudes at J, H, K and bolometric wavebands.
All magnitudes in Tables 1 and 2 have been corrected for extinction
following F89, assuming $A_J=0.06$, $A_H=0.03$ and $A_K=0.02$. We adopted as 
the maximum and minimum
magnitude simply the maximum and minimum
{\em observed\/} value (or interpolated value in the case of bolometric 
magnitude) reported in Glass et al (1990). Mean magnitudes 
were calculated as the average of the maximum and minimum 
observed (interpolated) magnitudes, which was also the definition of mean
magnitude adopted for the relations derived in F89. Multi-epoch observations
were available in Glass et al. (1990) for 48 of the 49 Miras studied in F89;
in the case of the star `GR13' only the mean magnitudes published in F89
were available.

\begin{table*}
\caption[]{O Miras: periods and J, H, K and bolometric magnitudes,
corrected for extinction, at mean, maximum and minimum light}
\begin{tabular}{lrrrrrrrrrrrrr}
\hline\noalign{\smallskip}
    &Period& \multicolumn{3}{c}{J band} &
\multicolumn{3}{c}{H band} &
\multicolumn{3}{c}{K band} &
\multicolumn{3}{c}{m$_{\rm{bol}}$}\\
Star&(days)&mean&max&min&mean&max&min&mean&max&min&mean&max&min\\
\noalign{\smallskip}
\hline\noalign{\smallskip}
%
%
w132      &155&12.66&12.40&12.91&11.85&11.52&12.19&11.65&11.37&11.92&14.48&14.20&14.75 \\
w151      &172&12.84&12.46&13.17&11.99&11.65&12.33&11.72&11.37&12.07&14.67&14.31&14.98 \\
w148      &185&12.96&12.60&13.35&12.12&11.74&12.50&11.80&11.38&12.22&14.74&14.41&15.16 \\
w158      &185&12.90&12.61&13.11&11.97&11.76&12.29&11.74&11.51&11.98&14.75&14.42&14.90 \\
w19       &189&12.60&12.39&12.82&11.73&11.52&11.93&11.49&11.32&11.66&14.33&14.19&14.62 \\
w77       &217&12.38&12.19&12.66&11.51&11.35&11.66&11.23&11.09&11.36&14.20&14.00&14.40 \\
w94       &220&12.44&12.29&12.59&11.55&11.42&11.67&11.27&11.15&11.39&14.23&14.09&14.37 \\
w74       &227&12.74&12.50&13.00&11.79&11.54&12.04&11.48&11.24&11.71&14.53&14.27&14.76 \\
w1        &233&12.80&12.34&13.54&11.90&11.50&12.79&11.62&11.12&12.13&14.55&14.16&15.39 \\
w140      &244&12.38&11.94&12.81&11.39&10.99&11.78&11.17&10.71&11.63&14.17&13.71&14.54 \\
w48       &279&12.12&11.85&12.39&11.25&10.95&11.53&10.97&10.68&11.26&13.97&13.64&14.19 \\
517-6551  &117&13.32&13.03&13.60&12.53&12.29&12.76&12.23&12.02&12.44&15.14&14.86&15.42 \\
512-6559  &141&13.31&12.83&13.79&12.49&12.04&12.94&12.11&11.70&12.51&15.05&14.66&15.61 \\
526-6754  &160&12.80&12.46&13.14&11.95&11.64&12.31&11.77&11.43&12.10&14.64&14.27&14.94 \\
528-6531  &195&12.46&12.18&12.73&11.62&11.37&11.95&11.45&11.15&11.76&14.31&13.99&14.54 \\
507-6639  &208&12.72&12.37&13.06&11.87&11.52&12.22&11.54&11.20&11.89&14.51&14.19&14.88 \\
533-6807  &247&12.51&12.02&13.00&11.73&11.24&12.20&11.36&10.93&11.79&14.28&13.85&14.84 \\
524-6543  &312&11.88&11.46&12.31&11.05&10.58&11.50&10.69&10.29&11.09&13.71&13.27&14.15 \\
505-6657  &311&11.92&11.53&12.31&11.07&10.72&11.41&10.65&10.29&11.01&13.73&13.36&14.11 \\
w126      &323&12.22&11.67&12.76&11.31&10.76&11.86&10.87&10.42&11.31&13.91&13.46&14.54 \\
c38       &128&13.08&12.73&13.47&12.37&12.08&12.65&12.10&11.78&12.42&14.87&14.58&15.28 \\
c11       &202&12.69&12.24&13.14&11.84&11.40&12.27&11.49&11.15&11.82&14.52&14.05&14.95 \\
c20       &210&12.88&12.37&13.40&12.01&11.49&12.31&11.52&11.08&11.96&14.56&14.17&15.18 \\
r120      &217&12.50&12.15&12.85&11.61&11.28&11.97&11.36&11.06&11.67&14.27&13.95&14.61 \\
r141      &255&12.13&11.73&12.53&11.27&10.84&11.65&10.97&10.63&11.31&13.87&13.54&14.36 \\
r110      &261&12.64&11.95&13.32&11.78&11.10&12.45&11.27&10.72&11.81&14.23&13.77&15.10 \\
r105      &420&11.58&11.22&11.93&10.64&10.29&10.98&10.27& 9.95&10.59&13.40&13.01&13.66 \\
537-6607  &284&12.26&11.85&12.68&11.39&10.95&11.81&11.00&10.60&11.39&13.99&13.64&14.49 \\
gr13      &202&12.68&     &     &11.83&     &     &11.57&     &     &14.50&     & \\
\noalign{\smallskip}
\hline
\end{tabular}
\end{table*}

\begin{table*}
\caption[]{C Miras: periods and J, H, K and bolometric magnitudes,
corrected for extinction, at mean, maximum and minimum light}
\begin{tabular}{lrrrrrrrrrrrrr}
\hline\noalign{\smallskip}
    &Period& \multicolumn{3}{c}{J band} &
\multicolumn{3}{c}{H band} &
\multicolumn{3}{c}{K band} &
\multicolumn{3}{c}{m$_{\rm{bol}}$}\\
Star&(days)&mean&max&min&mean&max&min&mean&max&min&mean&max&min\\
\noalign{\smallskip}
\hline\noalign{\smallskip}
%
%
w220      &286&12.70&12.06&13.33&11.47&11.03&11.92&10.81&10.57&11.04&14.12&13.77&14.48 \\
w46       &286&12.63&12.38&12.88&11.52&11.32&11.72&10.98&10.78&11.00&14.20&14.05&14.39 \\
w103      &351&13.64&12.95&14.70&12.00&11.47&12.53&10.76&10.35&11.17&14.14&13.81&14.52 \\
w30       &400&12.28&12.10&12.47&11.07&10.92&11.23&10.46&10.34&10.57&13.73&13.66&13.93 \\
530-6437  &157&13.06&12.42&13.70&12.39&11.77&13.00&12.06&11.52&12.60&14.71&14.25&15.57 \\
515-6617  &211&13.20&12.48&13.91&12.04&11.42&12.66&11.14&10.74&11.53&14.56&14.09&15.01 \\
528-6520  &231&12.65&12.06&13.24&11.63&11.19&12.06&11.06&10.82&11.29&14.29&13.87&14.70 \\
529-6759  &274&12.64&12.24&13.09&11.53&11.20&11.86&10.89&10.70&11.09&14.25&13.94&14.49 \\
515-6451  &284&12.90&12.41&13.43&11.67&11.31&12.04&10.79&10.54&11.03&14.21&13.95&14.49 \\
514-6605  &305&12.70&12.34&13.06&11.43&11.18&11.69&10.62&10.46&10.78&14.02&13.87&14.24 \\
502-6711  &308&12.58&11.91&13.25&11.35&10.90&11.81&10.51&10.26&10.76&13.92&13.58&14.22 \\
534-6531  &312&13.58&12.69&14.46&12.15&11.47&12.84&10.96&10.43&11.48&14.28&13.90&14.85 \\
529-6739  &319&12.92&12.29&13.46&11.59&11.10&12.07&10.58&10.25&10.91&13.91&13.69&14.37 \\
541-6631  &328&13.16&12.38&14.73&11.91&11.13&12.72&10.48&10.12&11.27&14.09&13.59&14.55 \\
515-6438  &365&13.06&12.45&13.98&11.88&11.15&12.56&10.88&10.38&11.38&14.27&13.80&14.91 \\
537-6740  &418&12.52&12.23&12.80&11.25&11.06&11.45&10.45&10.33&10.57&13.90&13.72&14.02 \\
c7        &326&12.58&12.02&12.98&11.36&11.01&11.66&10.67&10.50&10.83&13.99&13.75&14.27 \\
r153      &370&13.10&12.39&13.82&11.58&11.10&12.06&10.50&10.23&10.81&14.03&13.67&14.22 \\
519-6454  &242&12.79&12.31&13.27&11.67&11.31&12.01&11.07&10.84&11.29&14.33&14.03&14.70 \\
520-6528  &234&12.46&12.36&12.56&11.57&11.35&11.73&11.26&11.16&11.36&14.17&14.02&14.43 \\
\noalign{\smallskip}
\hline
\end{tabular}
\end{table*}

Using these data we carried out linear regression fits to PL relations of the
form
\begin{equation}
m = a + b \log P
\label{eq:eq4}
\end{equation}
\begin{equation}
m_{\rm{max}} = a + b \log P
\label{eq:eq5}
\end{equation}
and
\begin{equation}
m_{\rm{min}} = a + b \log P
\label{eq:eq6}
\end{equation}
where $m$ denotes apparent J, H, K and bolometric magnitude, corrected
for extinction, as appropriate and $a$ and $b$ are constants.

As will be clear from Figures 1 to 6 in Section 4, in most cases the
existence of a tightly correlated PL relation
at mean and maximum light was immediately evident
from a scatterplot of apparent magnitude against log period. Notwithstanding
this, we considered it instructive -- particularly for the more marginal
PL relations -- to determine quantitatively the statistical significance
of including a log period term in each of our regression fits. In order to do
this we applied the same statistical test which was introduced in KH, and
which is described in detail in the appendix of that paper, involving the
partial multiple correlation coefficient, $\rho$, of the regression
(c.f. Graybill 1976). If $\rho$ equals zero then the log period term makes no
contribution to a reduction in the dispersion of the fit and is
effectively redundant.
For each regression we computed the sample value of $\rho$,
denoted by $\hat{\rho}$. Under the null hypothesis that the true value of
$\rho$ is equal to zero then $\hat{\rho}^2$ has an F distribution (c.f. KH).

We also carried out fits to PLC relations of the form
\begin{equation}
m = a + b \log P + c (J-K)
\label{eq:eq7}
\end{equation}
\begin{equation}
m_{\rm{max}} = a + b \log P + c (J-K)_{\rm{max}}
\label{eq:eq8}
\end{equation}
and
\begin{equation}
m_{\rm{min}} = a + b \log P + (J-K)_{\rm{min}}
\label{eq:eq9}
\end{equation}
where $(J-K)$ denotes dereddened colour, and also to the corresponding 
equations for $(J-H)$ colour. 
We defined maximum and minimum colour as the colour at the phase at which 
respectively the maximum and minimum magnitude was observed. We applied
the same statistical test as for the PL relations to determine the
significance of adding the colour term in each PLC relation.

We compared the fitted relations obtained using mean, maximum and minimum
magnitudes as defined above with those derived
using magnitudes calculated from a first order fourier
fit to the light curve of each Mira. In all cases we found no
significant difference in the slopes, zero points and dispersions of the
fitted relations at either maximum or mean light. The same conclusion
regarding the robustness of mean magnitudes was reached in F89, where
mean values obtained from averaging the maximum and minimum magnitudes
were compared with the average of maximum and minimum intensities and
also with the results of fourier fits to both magnitudes and intensities. The
robustness of mean and maximum PL relations to the choice of definition
for mean and maximum light is in complete accordance with the
results of Hendry, Kanbur \& Clarke (1997, in prep.), in which we investigate
the statistical properties of various different estimators of mean and maximum
light -- including those adopted here and those derived from fitting low order
fourier series -- as a function of number of sampled phase points, phase
coverage, light curve shape and limiting magnitude.

\section{PL Relation Results}

The results of our regression fits to equations 4, 5 and 6 are presented in
Tables 3 to 6 and illustrated in Figures 1 to 6. Column 1 in each Table
lists the type of regression fit -- i.e. to mean, maximum or minimum apparent
magnitude. Column 2 indicates the type of Mira sample used: C Miras only,
O Miras only or both Mira types (denoted `O+C'), and column 3 gives the
number, $n$, of Miras in each sample. Note that for the mean relations we
used the full sample of 29 O Miras, identical to that used in F89, while
for the minimum and maximum light relations we used the sample of 28 O Miras 
for which phase information was available. Columns 4 to 7 give the 
fitted values of the
zero point, $a$, and slope, $b$, of the relations with their associated
standard errors, $\sigma_a$ and $\sigma_b$. Column 8 indicates the dispersion,
$\sigma$ (in magnitudes) of the regression fit and column 9 gives the
percentage root mean square error, $\Delta$, of the corresponding distance
indicator which one would derive from the PL relation, i.e.
$\Delta \simeq 46.1 \sigma \%$. Finally, column 10 gives the value of
$\hat{\rho}$, the partial multiple correlation coefficient computed for
the sample data and column 11 indicates the probability
(denoted by `Prob') that
$\hat{\rho}^2$ be equal to (or greater than) its computed value under the
null hypothesis that the true value of $\rho$ is equal to zero.

Figures 1 and 2 show scatterplots of mean and maximum magnitude respectively
against log period for the oxygen-rich Miras. The fitted regression lines
in each waveband are also drawn on the plots. Figures 3 and 4 show the
corresponding scatterplots for the carbon Miras in our sample and
Figures 5 and 6 show the scatterplots for the composite sample of 
C and O Miras.

\begin{figure*}
\centerline{\psfig{file=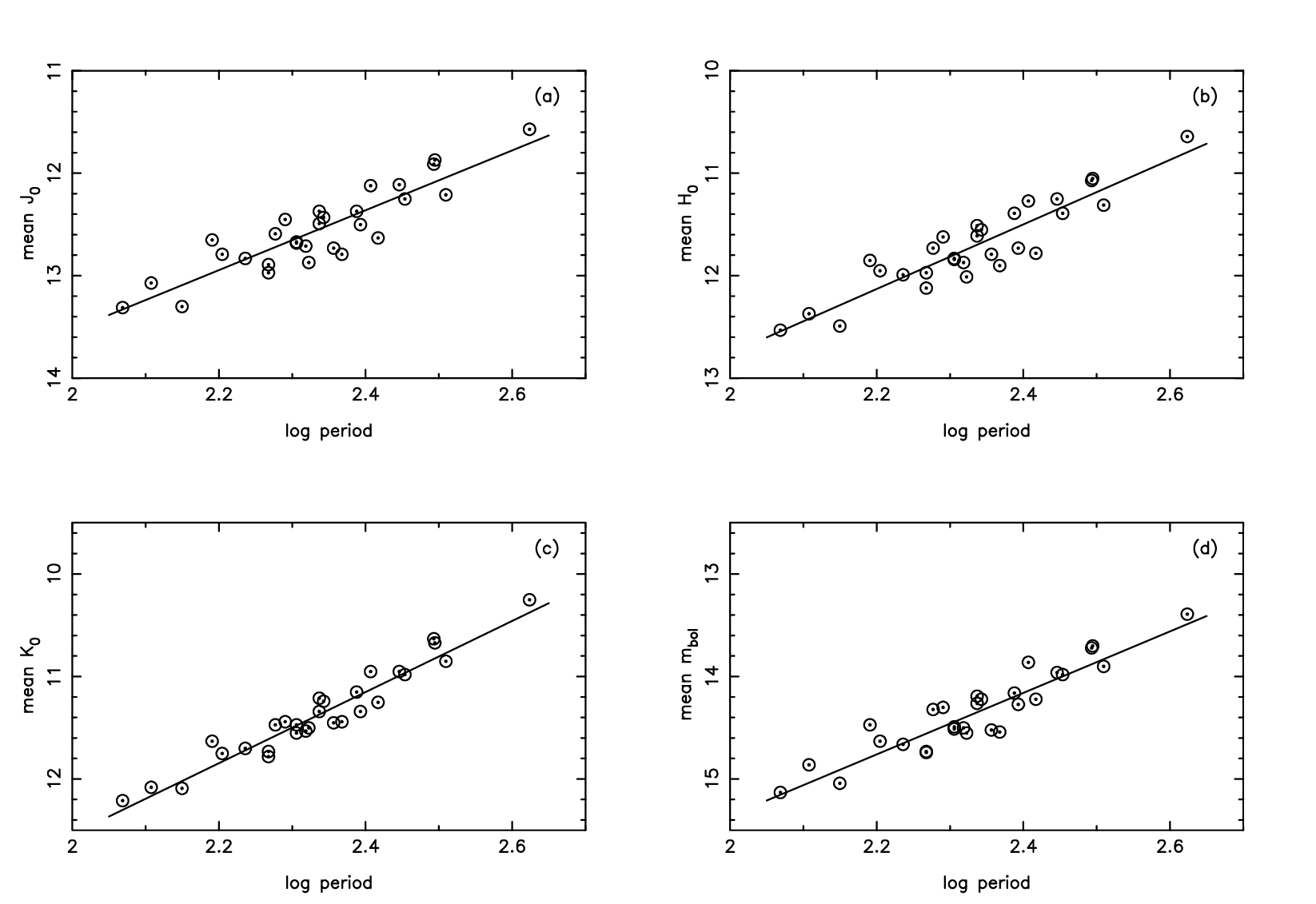,width=15cm,height=10cm,angle=-90}}
\caption[]{Fitted PL relations at mean light, for (a) J band, (b) H band,
(c) K band, and (d) bolometric apparent magnitudes,
derived for the sample of 29 oxygen-rich Miras.}
\label{fig:fig1}
\end{figure*}

\begin{figure*}
\centerline{\psfig{file=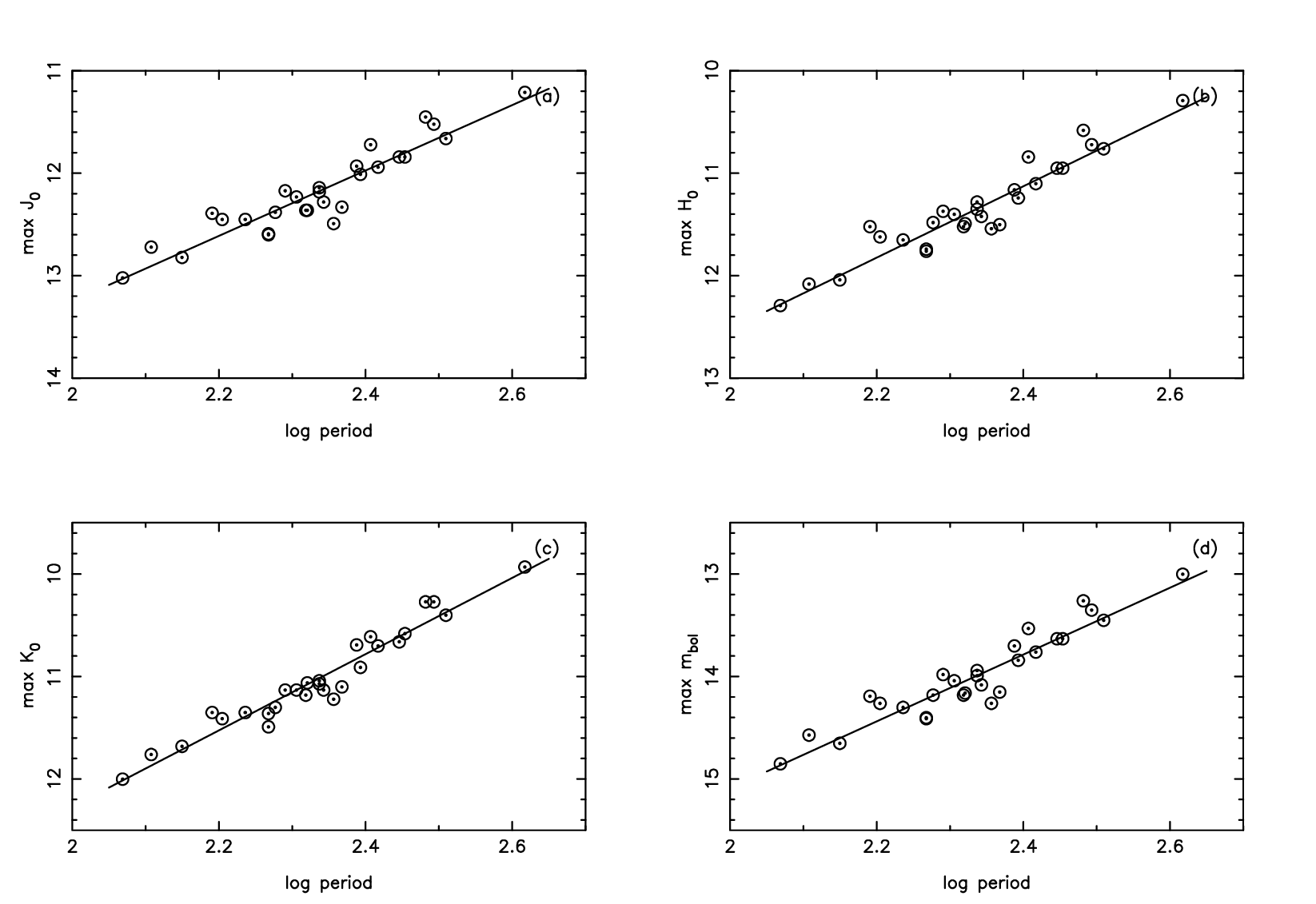,width=15cm,height=10cm,angle=-90}}
\caption[]{Fitted PL relations at maximum light, for (a) J band, (b) H band,
(c) K band, and (d) bolometric apparent magnitudes,
derived for the sample of 28 oxygen-rich Miras.}
\label{fig:fig2}
\end{figure*}

\begin{figure*}
\centerline{\psfig{file=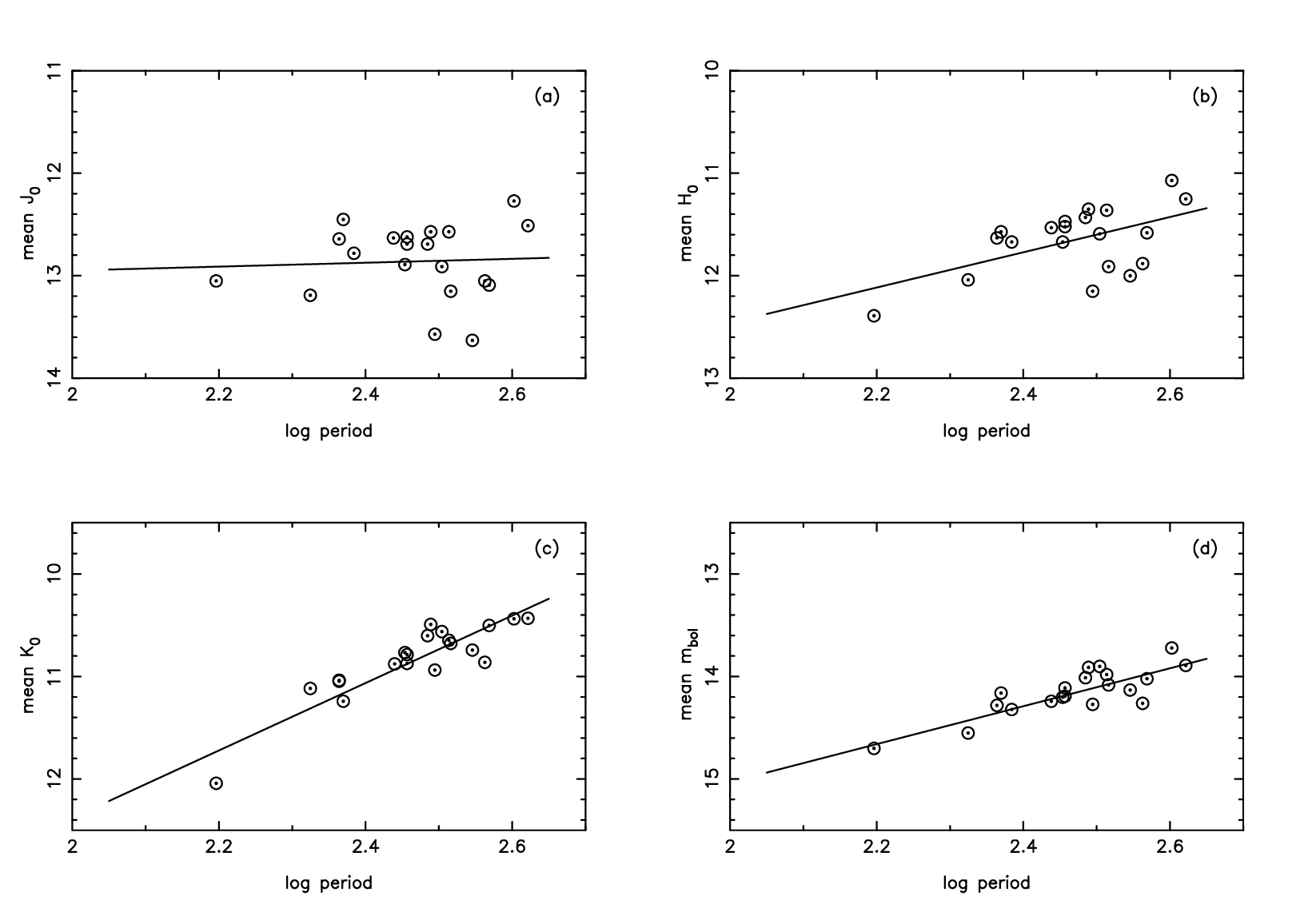,width=15cm,height=10cm,angle=-90}}
\caption[]{Fitted PL relations at mean light, for (a) J band, (b) H band,
(c) K band, and (d) bolometric apparent magnitudes,
derived for the sample of 20 carbon-rich Miras.}
\label{fig:fig3}
\end{figure*}

\begin{figure*}
\centerline{\psfig{file=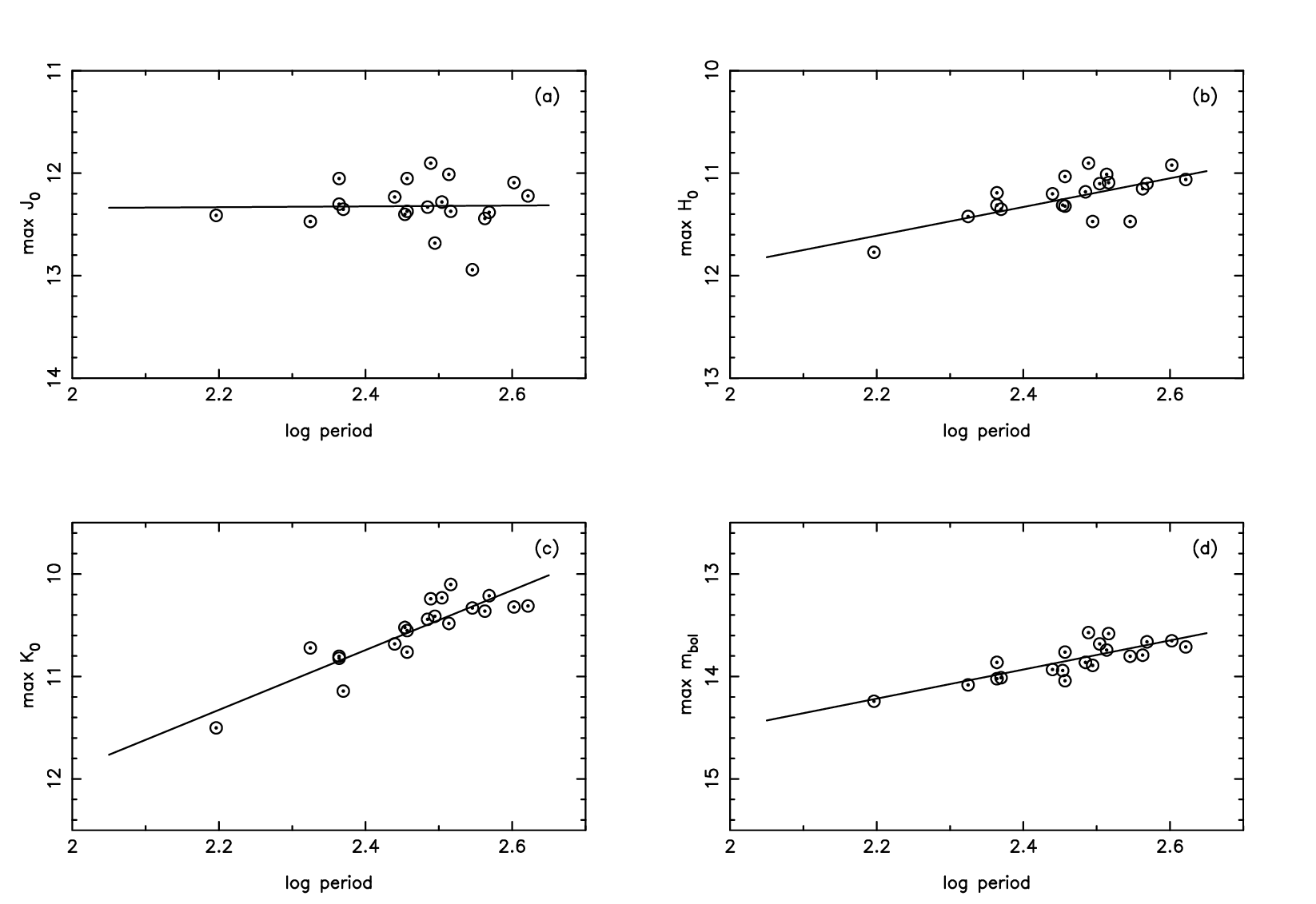,width=15cm,height=10cm,angle=-90}}
\caption[]{Fitted PL relations at maximum light, for (a) J band, (b) H band,
(c) K band, and (d) bolometric apparent magnitudes,
derived for the sample of 20 carbon-rich Miras.}
\label{fig:fig4}
\end{figure*}

\begin{figure*}
\centerline{\psfig{file=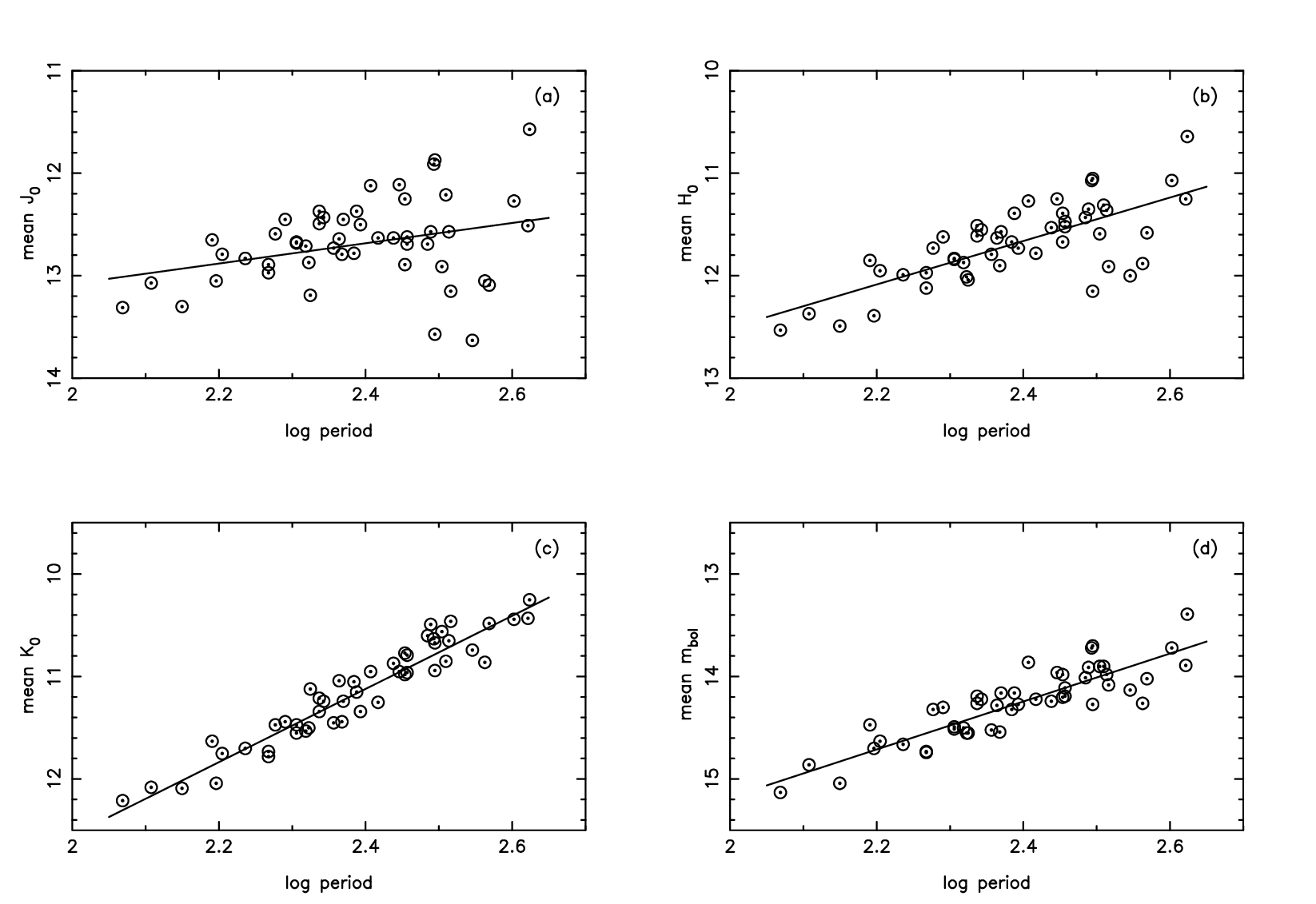,width=15cm,height=10cm,angle=-90}}
\caption[]{Fitted PL relations at mean light, for (a) J band, (b) H band,
(c) K band, and (d) bolometric apparent magnitudes,
derived for the sample of 49 carbon and oxygen-rich Miras.}
\label{fig:fig5}
\end{figure*}

\begin{figure*}
\centerline{\psfig{file=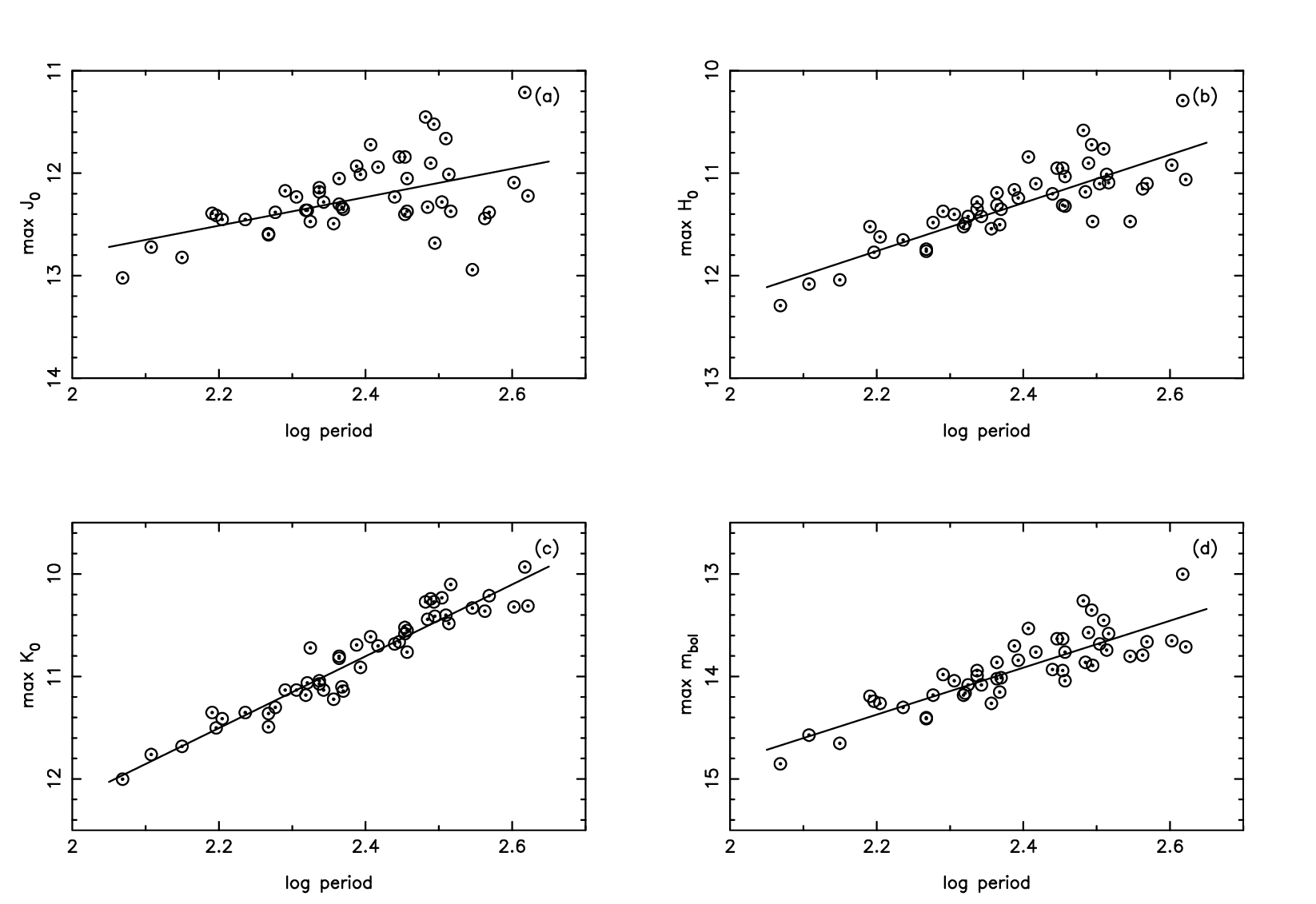,width=15cm,height=10cm,angle=-90}}
\caption[]{Fitted PL relations at maximum light, for (a) J band, (b) H band,
(c) K band, and (d) bolometric apparent magnitudes,
derived for the sample of 48 carbon and oxygen-rich Miras.}
\label{fig:fig6}
\end{figure*}

\begin{table*}
\caption[]{J band fitted PL relations}
\noindent
\begin{tabular}[t]{ccccccccccl}
fit &  sample & $n$ & $a$ & $\sigma_a$ & $b$ & $\sigma_b$ & $\sigma$
& $\Delta$ & $\hat{\rho}$ & Prob\\ \hline
& & & & & & & & & & \\
mean & C   & 20 & 13.33 & 2.05 & -0.19 & 0.83 & 0.368 & 16.9 &
 -0.054 & 0.820\\
& & & & & & & & & & \\
 max & C   & 20 & 12.42 & 1.34 & -0.04 & 0.54 & 0.243 & 11.2 & 
 -0.017 & 0.942\\
& & & & & & & & & & \\
 min & C   & 20 & 13.18 & 3.68 &  0.11 & 1.49 & 0.668 & 30.8 &
  0.02  & 0.940\\
& & & & & & & & & & \\
mean & O   & 29 & 19.37 & 0.67 & -2.92 & 0.28 & 0.187 &  8.6 &
 -0.891 & $8.43 \times 10^{-11}$\\
& & & & & & & & & & \\
 max & O   & 28 & 19.63 & 0.57 & -3.19 & 0.24 & 0.158 &  7.3 &
 -0.932 & $5.46 \times 10^{-13}$\\
& & & & & & & & & & \\
 min & O   & 28 & 19.12 & 1.05 & -2.65 & 0.45 & 0.292 & 13.4 &
 -0.760    & $3.13 \times 10^{-6}$ \\
& & & & & & & & & & \\
mean & O+C & 49 & 15.06 & 1.02 & -0.99 & 0.43 & 0.392 & 18.1 &
 -0.321 & 0.020\\
& & & & & & & & & & \\
 max & O+C & 48 & 15.57 & 0.83 & -1.39 & 0.35 & 0.317 & 14.6 &
 -0.508 & $2.27 \times 10^{-4}$\\
& & & & & & & & & & \\
 min & O+C & 48 & 14.03 & 1.56 & -0.37 & 0.65 & 0.593 & 27.3 &
  -0.083  & 0.577      \\
\end{tabular}
\end{table*}

\begin{table*}
\caption[]{H band fitted PL relations}
\noindent
\begin{tabular}[t]{ccccccccccl}
fit &  sample & $n$ & $a$ & $\sigma_a$ & $b$ & $\sigma_b$ & $\sigma$
& $\Delta$ & $\hat{\rho}$ & Prob\\ \hline
& & & & & & & & & & \\
mean & C   & 20 & 15.90 & 1.56 & -1.72 & 0.63 & 0.280 & 12.9 &
 -0.541 & $1.38 \times 10^{-2}$\\
& & & & & & & & & & \\
 max & C   & 20 & 14.69 & 0.89 & -1.40 & 0.36 & 0.161 &  7.4 & 
 -0.678 & $1.02 \times 10^{-3}$\\
& & & & & & & & & & \\
 min & C   & 20 & 16.96 & 2.48 & -1.97 & 1.01 & 0.451 & 20.7 &
  -0.42 & $6.55 \times 10^{-2}$ \\
& & & & & & & & & & \\
mean & O   & 29 & 19.06 & 0.64 & -3.15 & 0.27 & 0.180 &  8.3 &
 -0.912 & $6.31 \times 10^{-12}$\\
& & & & & & & & & & \\
 max & O   & 28 & 19.48 & 0.53 & -3.48 & 0.23 & 0.147 &  6.8 &
 -0.949 & $1.37 \times 10^{-14}$\\
& & & & & & & & & & \\
 min & O   & 28 & 18.71 & 1.07 & -2.84 & 0.46 & 0.299 & 13.8 &
 -0.771 & $1.56 \times 10^{-6}$ \\
& & & & & & & & & & \\
mean & O+C & 49 & 16.75 & 0.70 & -2.12 & 0.29 & 0.269 & 12.4 &
 -0.724 & $4.01 \times 10^{-9}$\\
& & & & & & & & & & \\
 max & O+C & 48 & 16.93 & 0.57 & -2.35 & 0.24 & 0.216 & 9.9 &
 -0.823 & $5.18 \times 10^{-13}$\\
& & & & & & & & & & \\
 min & O+C & 48 & 16.66 & 1.03 & -1.91 & 0.43 & 0.394 & 18.1 &
 -0.546 &$5.89 \times 10^{-5}$  \\
\end{tabular}
\end{table*}

\begin{table*}
\caption[]{K band fitted PL relations}
\noindent
\begin{tabular}[t]{ccccccccccl}
fit &  sample & $n$ & $a$ & $\sigma_a$ & $b$ & $\sigma_b$ & $\sigma$
& $\Delta$ & $\hat{\rho}$ & Prob\\ \hline
& & & & & & & & & & \\
mean & C   & 20 & 18.96 & 0.98 & -3.29 & 0.40 & 0.176 & 8.1 &
 -0.890 & $1.45 \times 10^{-7}$\\
& & & & & & & & & & \\
 max & C   & 20 & 17.75 & 0.93 & -2.92 & 0.38 & 0.169 &  7.8 & 
 -0.877 & $3.97 \times 10^{-7}$\\
& & & & & & & & & & \\
 min & C   & 20 & 19.64 & 1.54 & -3.45 & 0.62 & 0.279 & 12.8 &
 -0.794 & $2.89 \times 10^{-5}$  \\
& & & & & & & & & & \\
mean & O   & 29 & 19.48 & 0.45 & -3.47 & 0.19 & 0.126 &  5.8 &
 -0.961 & $1.29 \times 10^{-16}$\\
& & & & & & & & & & \\
 max & O   & 28 & 19.71 & 0.45 & -3.72 & 0.19 & 0.124 &  5.7 &
 -0.967 & $5.28 \times 10^{-17}$\\
& & & & & & & & & & \\
 min & O   & 28 & 19.25 & 0.77 & -3.23 & 0.33 & 0.215 &  9.9 &
 -0.886 & $3.48 \times 10^{-10}$ \\
& & & & & & & & & & \\
mean & O+C & 49 & 19.69 & 0.39 & -3.57 & 0.16 & 0.150 & 6.9 &
 -0.954 & $2.86 \times 10^{-26}$\\
& & & & & & & & & & \\
 max & O+C & 48 & 19.21 & 0.39 & -3.50 & 0.16 & 0.148 & 6.8 &
 -0.953 & $1.15 \times 10^{-25}$\\
& & & & & & & & & & \\
 min & O+C & 48 & 20.00 & 0.65 & -3.57 & 0.27 & 0.247 & 11.4 &
 -0.889 &$3.06 \times 10^{-17}$ \\
\end{tabular}
\end{table*}

\begin{table*}
\caption[]{$m_{\rm{bol}}$ fitted PL relations}
\noindent
\begin{tabular}[t]{ccccccccccl}
fit &  sample & $n$ & $a$ & $\sigma_a$ & $b$ & $\sigma_b$ & $\sigma$
& $\Delta$ & $\hat{\rho}$ & Prob\\ \hline
& & & & & & & & & & \\
mean & C   & 20 & 18.73 & 0.74 & -1.85 & 0.30 & 0.132 & 6.1 &
 -0.826 & $7.21 \times 10^{-6}$\\
& & & & & & & & & & \\
 max & C   & 20 & 17.34 & 0.59 & -1.42 & 0.24 & 0.107 & 5.0 & 
 -0.937 & $1.85 \times 10^{-13}$\\
& & & & & & & & & & \\
 min & C   & 20 & 21.25 & 1.39 & -2.73 & 0.56 & 0.252 & 11.6 &
 -0.753 & $1.29 \times 10^{-4}$  \\
& & & & & & & & & & \\
mean & O   & 29 & 21.36 & 0.57 & -3.00 & 0.24 & 0.160 &  7.4 &
 -0.920 & $1.26 \times 10^{-12}$\\
& & & & & & & & & & \\
 max & O   & 28 & 21.61 & 0.55 & -3.26 & 0.24 & 0.154 &  7.1 &
 -0.937 & $1.85 \times 10^{-13}$\\
& & & & & & & & & & \\
 min & O   & 28 & 21.11 & 1.07 & -2.73 & 0.46 & 0.297 & 13.7 &
 0.760 & $2.62 \times 10^{-6}$  \\
& & & & & & & & & & \\
mean & O+C & 49 & 19.86 & 0.45 & -2.34 & 0.19 & 0.173 & 8.0 &
 -0.875 & $2.11 \times 10^{-16}$\\
& & & & & & & & & & \\
 max & O+C & 48 & 19.41 & 0.50 & -2.29 & 0.21 & 0.190 & 8.7 &
 -0.851 & $1.85 \times 10^{-14}$\\
& & & & & & & & & & \\
 min & O+C & 48 & 20.55 & 0.74 & -2.47 & 0.31 & 0.280 & 12.9 &
 -0.76 & $2.74 \times 10^{-10}$  \\
\end{tabular}
\end{table*}

It is evident from Tables 3 to 6 and from Figures 2, 4 and 6 that statistically
significant PL relations clearly exist at maximum light in almost all of the
cases considered  -- a fact which is supported quantitatively by the values of
the partial multiple correlation coefficient, which are generally different
from zero at a very high level of significance. The only clear exception to
this trend is the case of the J band maximum light PL relation for C Miras,
which we discuss further below.

For the O Miras the K band PL relation has the smallest dispersion at 
maximum light -- about 20\% smaller than for the relations in the other
wavebands. The maximum light relation for the composite sample of both types of 
Miras also has the smallest dispersion in the K band. Similar behaviour was 
found in F89 for the mean light relations, which we also confirm here.

It is also clear from Tables 3 to 6 that the dispersion of the minimum light
PL relations is considerably larger than that of both the mean light and
maximum light relations in all cases. We comment on this in section 5 below.

Comparing all of our results for mean light relations with those of F89,
we see that our fitted coefficients, standard errors and dispersions
are in excellent agreement in all cases.

We can see from Tables 3 to 6 that the dispersions of the maximum light
PL relations are smaller than those of the corresponding mean light
relations in every case considered, with the sole exception of the `O+C'
relation for bolometric magnitudes. Note also that the standard errors
of the fitted regression coefficients are also consistently smaller for the
maximum light relations. To assess the statistical significance of this
result requires some care, however. We cannot simply apply a standard
ratio-of-variance F test (c.f. Graybill 1976) to the data since such
a test assumes that the variances are statistically independent. This
condition is clearly not satisfied here, as the residuals of our maximum
and mean light PL relations are likely to be highly correlated. A failure
to account for this correlation would result in underestimating the
significance of the reduction in dispersion. We tackle this
problem numerically, first computing the correlation coefficient
of the mean and maximum light residuals and then -- with this correlation
coefficient -- generating a large number of Monte Carlo simulations to
estimate the probability
density function of the sample ratio of mean to maximum light dispersion,
under the null hypothesis that these dispersions are equal. We then
determine the statistical significance of the observed reduction in
dispersion by considering the the extent of the tail of our estimated
probablility density function in the standard manner.

The results of applying this significance test are given in Table 7.
The first and second columns indicate the waveband and type of Mira
sample under consideration, the third column indicates the ratio,
$R$, of the variance at mean light to the variance at maximum light.
The fourth column gives $r$, the sample correlation coefficient of the
residuals at mean and maximum light, and the final column
indicates the probability (denoted by `Prob') of obtaining as large
(or larger) a value of $R$ under the null hypothesis that the true value of
the ratio is equal to unity -- i.e. the dispersion at mean and maximum
light is identical -- and the true correlation coefficient of the mean and
maximum light residuals is equal to $r$.

\begin{table}
\caption[]{Significance of the reduction in dispersion of maximum light PL
relations compared with mean light PL relations. Column headings are
explained in the text}
\noindent
\begin{tabular}[t]{ccccl}
      waveband & sample & $R$ & $r$ & Prob\\ \hline
               & & & & \\
             J & C  & 2.293 & 0.827 & $1.4 \times 10^{-3}$\\
             H & C  & 3.025 & 0.748 & $6.0 \times 10^{-4}$\\
             K & C  & 1.084 & 0.701 & 0.398\\
$m_{\rm{bol}}$ & C  & 1.494 & 0.513 & 0.160\\
               & & & & \\
             J & O  & 1.401 & 0.795 & $7.8 \times 10^{-2}$\\
             H & O  & 1.500 & 0.753 & $5.8 \times 10^{-2}$\\
             K & O  & 1.032 & 0.688 & 0.454\\
$m_{\rm{bol}}$ & O  & 1.079 & 0.858 & 0.348\\
               & & & & \\
             J & O+C & 1.529 & 0.904 & $5.0 \times 10^{-4}$\\
             H & O+C & 1.551 & 0.818 & $4.4 \times 10^{-3}$\\
             K & O+C & 1.027 & 0.677 & 0.451\\
$m_{\rm{bol}}$ & O+C & 0.829 & 0.830 & 0.875\\
\end{tabular}
\end{table}

We can see from Table 7 that the ratio, $R$, is greater than unity in
all cases except that of the `O+C' relation for bolometric magnitudes.
The reduction in scatter is least significant for the K band
Miras -- i.e. the tight PL relation already displayed at mean light is
not improved as much by the use of maximum light as in the other
wavebands -- but is still non-negligible. A significant reduction
is seen for both the H and J band relations. Although the dispersion at
maximum light is slightly larger than at mean light for the `O+C' bolometric
relation, the increase in dispersion is not statistically significant.
Table 7 illustrates the importance of
accounting for the correlation between the mean and maximum light
residuals: the reduction in dispersion for the C Miras in the J band is
marginally smaller than that in the H band, but is marginally more
significant because the J band residuals are more highly correlated.

As mentioned above, the J band PL relation for C Miras at maximum light is
seen from Figure 4 to be essentially flat. This is confirmed in Table 3,
where we see that the fitted coefficient of log period is only -0.02,
consistent with zero, and the partial multiple correlation coefficient
is not significantly different from zero. A similarly flat relation is
seen in Figure 3, for the J band relation at mean light -- as previously
reported by F89. The existence of an H, K and bolometric magnitude PL
relation, at both mean and maximum light, for the C Miras is somewhat
more convincing in Figures 3 and 4: all have significantly non-zero
partial multiple correlation coefficients and regression coefficients
of log period. If we consider only those C Miras with periods greater than
250 days, however, then the H, K and bolometric relations are considerably
flatter -- i.e. the longer period Miras are more consistent with having
constant absolute magnitude. We comment further on this in the next section.

Finally, F89 found that the slope of the bolometric magnitude PL relation
at mean light was shallower for C Miras than for O Miras. Our results
confirm this conclusion and indicate that it is also true -- and indeed
is considerably more pronounced -- at maximum light.

\section{PLC Results}

F89 also presented evidence for the existence of PLC relations at mean light,
for both the carbon and oxygen Miras based on $(J-K)$ colours. By considering
the correlation of the PL relation residuals with $(J-K)$, F89 found evidence
of a significant mean colour term for the O Miras at all wavelengths, but for
the C Miras the colour term was highly significant only for the J band PLC
relation, and was found to be marginal for the K band and bolometric relations.
F89 also showed that, where significant, the colour term was intrinsic and
could not be attributed to differential reddening. In this paper we have 
derived PLC relations at mean, maximum and minimum light using both $(J-K)$
and $(J-H)$ colours. We present the results of our regression fits to equations
7 to 9 in Tables 8 to 10 below, with the
corresponding results for $(J-H)$ colours in Tables 11 to 13.
The columns of these Tables are as in Tables 3 to 6, with two additional
columns giving the fitted value and standard error of the colour
coefficient, $c$. Note that in the case of the PLC relations with
$(J-K)$ colours we do not present the K band results since it is
straightforward to show that these are trivially related to those at
J band: i.e. $a_K = a_J$, $b_K = b_J$ and $c_K = c_J - 1$. Moreover, one may
also show that the dispersions of the J and K band PLC relations, and
the standard error of the coefficients, are identical. The J and H band
PLC relations based on $(J-H)$ colour are similarly related in a trivial
way.

\begin{table*}
\caption[]{J band fitted PLC relations using $(J-K)$ colour}
\noindent
\begin{tabular}[t]{ccccccccccccl}
fit &  sample & $n$ & $a$ & $\sigma_a$ & $b$ & $\sigma_b$ & $c$
& $\sigma_c$ & $\sigma$ & $\Delta$ & $\hat{\rho}$ & Prob\\ \hline
& & & & & & & & & & & & \\
mean & C   & 20 & 18.37 & 1.18 & -2.97 & 0.53 & 0.89 & 0.11 &
 0.177 & 8.2 & 0.884 & $5.14 \times 10^{-7}$\\
& & & & & & & & & & & & \\
 max & C   & 20 & 16.03 & 1.01 & -1.99 & 0.46 & 0.68 & 0.12 & 
 0.144 & 6.6 & 0.817 & $1.96 \times 10^{-5}$\\
& & & & & & & & & & & & \\
 min & C   & 20 & 20.88 & 1.68 & -4.14 & 0.74 & 1.19 &  0.12 &
 0.268 & 12.4 & 0.921 & $2.30 \times 10^{-8}$\\
& & & & & & & & & & & & \\
mean & O   & 29 & 19.57 & 0.37 & -3.98 & 0.21 & 1.91 &  0.24 &
 0.103 & 4.8 & 0.840 & $2.29 \times 10^{-8}$\\
& & & & & & & & & & & & \\
 max & O   & 28 & 19.76 & 0.43 & -4.04 & 0.27 & 1.61 &  0.36 &
 0.120 & 5.5 & 0.665 & $1.55 \times 10^{-4}$\\
& & & & & & & & & & & & \\
 min & O   & 28 & 19.37 & 0.69 & -3.77 & 0.35 & 1.91 &  0.32 &
 0.191 & 8.8 & 0.767 & $3.02 \times 10^{-6}$\\
& & & & & & & & & & & & \\
mean & O+C & 49 & 19.36 & 0.47 & -3.39 & 0.22 & 0.93 & 0.06 &
 0.149 & 6.9 & 0.927 & $3.35 \times 10^{-21}$\\
& & & & & & & & & & & & \\
 max & O+C & 48 & 18.97 & 0.47 & -3.37 & 0.22 & 0.94 & 0.07 &
 0.149 & 6.8 & 0.886 & $1.28 \times 10^{-16}$\\
& & & & & & & & & & & & \\
 min & O+C & 48 & 20.01 & 0.77 & -3.58 & 0.35 & 1.00 & 0.06 &
 0.249 & 11.5 & 0.910 & $9.04 \times 10^{-19}$\\
\end{tabular}
\end{table*}

\begin{table*}
\caption[]{H band fitted PLC relations using $(J-K)$ colour}
\noindent
\begin{tabular}[t]{ccccccccccccl}
fit &  sample & $n$ & $a$ & $\sigma_a$ & $b$ & $\sigma_b$ & $c$
& $\sigma_c$ & $\sigma$ & $\Delta$ & $\hat{\rho}$ & Prob\\ \hline
& & & & & & & & & & & & \\
mean & C   & 20 & 19.06 & 1.31 & -3.46 & 0.59 & 0.56 & 0.13 &
 0.197 & 9.1 & 0.729 & $4.00 \times 10^{-4}$\\
& & & & & & & & & & & & \\
 max & C   & 20 & 16.50 & 0.91 & -2.38 & 0.42 & 0.34 &  0.10 & 
 0.130 & 6.0 & 0.619 & $4.69 \times 10^{-3}$\\
& & & & & & & & & & & & \\
 min & C   & 20 & 21.37 & 1.80 & -4.41 & 0.80 & 0.68 &  0.13 &
 0.288 & 13.3 & 0.784 & $7.25 \times 10^{-5}$\\
& & & & & & & & & & & & \\
mean & O   & 29 & 19.24 & 0.40 & -4.11 & 0.22 & 1.74 &  0.26 &
 0.111 & 5.1 & 0.796 & $4.17 \times 10^{-7}$\\
& & & & & & & & & & & & \\
 max & O   & 28 & 19.59 & 0.42 & -4.23 & 0.26 & 1.42 &  0.35 &
 0.117 & 5.4 & 0.627 & $4.65 \times 10^{-4}$\\
& & & & & & & & & & & & \\
 min & O   & 28 & 18.93 & 0.83 & -3.81 & 0.42 & 1.67 &  0.39 &
 0.231 & 10.6 & 0.653 & $2.23 \times 10^{-4}$\\
& & & & & & & & & & & & \\
mean & O+C & 49 & 19.27 & 0.52 & -3.53 & 0.24 & 0.55 & 0.06 &
 0.165 & 7.6 & 0.794 & $1.62 \times 10^{-11}$\\
& & & & & & & & & & & & \\
 max & O+C & 48 & 18.95 & 0.44 & -3.53 & 0.21 & 0.56 & 0.07 &
 0.138 & 6.4 & 0.773 & $1.89 \times 10^{-10}$\\
& & & & & & & & & & & & \\
min & O+C & 48 & 19.76 & 0.87 & -3.58 & 0.39 & 0.52 & 0.08 &
 0.280 & 12.9 & 0.711& $2.14 \times 10^{-8}$\\
\end{tabular}
\end{table*}

\begin{table*}
\caption[]{$m_{\rm{bol}}$ fitted PLC relations using $(J-K)$ colour}
\noindent
\begin{tabular}[t]{ccccccccccccl}
fit &  sample & $n$ & $a$ & $\sigma_a$ & $b$ & $\sigma_b$ & $c$
& $\sigma_c$ & $\sigma$ & $\Delta$ & $\hat{\rho}$ & Prob\\ \hline
& & & & & & & & & & & & \\
mean & C   & 20 & 19.79 & 0.73 & -2.44 & 0.33 & 0.19 & 0.07 &
 0.112 & 5.2 & 0.445 & $5.65 \times 10^{-2}$\\
& & & & & & & & & & & & \\
 max & C   & 20 & 17.55 & 0.77 & -1.53 & 0.36 & 0.04 &  0.09 & 
 0.111 & 5.1 & 0.010 & 0.671\\
& & & & & & & & & & & & \\
 min & C   & 20 & 22.80 & 1.41 & -3.59 & 0.62 & 0.24 &  0.10 &
 0.226 & 10.4 & 0.493 & $3.21 \times 10^{-2}$\\
& & & & & & & & & & & & \\
mean & O   & 29 & 21.51 & 0.39 & -3.80 & 0.22 & 1.43 &  0.26 &
 0.111 & 5.1 & 0.736 & $7.96 \times 10^{-6}$\\
& & & & & & & & & & & & \\
 max & O   & 28 & 21.74 & 0.43 & -4.08 & 0.26 & 1.54 &  0.36 &
 0.119 & 5.5 & 0.651 & $2.36 \times 10^{-4}$\\
& & & & & & & & & & & & \\
 min & O   & 28 & 21.36 & 0.73 & -3.83 & 0.37 & 1.88 &  0.34 &
 0.204 & 9.4 & 0.741 & $1.00 \times 10^{-5}$\\
& & & & & & & & & & & & \\
mean & O+C & 49 & 21.13 & 0.43 & -3.05 & 0.20 & 0.27 & 0.05 &
 0.138 & 6.3 & 0.618 & $2.88 \times 10^{-6}$\\
& & & & & & & & & & & & \\
 max & O+C & 48 & 20.82 & 0.48 & -3.11 & 0.23 & 0.39 & 0.07 &
 0.151 & 7.0 & 0.616 & $4.04 \times 10^{-6}$\\
& & & & & & & & & & & & \\
 min & O+C & 48 & 21.85 & 0.80 & -3.17 & 0.36 & 0.22 & 0.07 &
 0.257 & 11.8 & 0.420 & $3.28 \times 10^{-3}$\\
\end{tabular}
\end{table*}

\begin{table*}
\caption[]{J band fitted PLC relations using $(J-H)$ colour}
\noindent
\begin{tabular}[t]{ccccccccccccl}
fit &  sample & $n$ & $a$ & $\sigma_a$ & $b$ & $\sigma_b$ & $c$
& $\sigma_c$ & $\sigma$ & $\Delta$ & $\hat{\rho}$ & Prob\\ \hline
& & & & & & & & & & & & \\
mean & C   & 20 & 18.35 & 1.78 & -3.17 & 0.86 & 1.95 & 0.42 &
 0.253 & 11.7 & 0.744 & $2.59 \times 10^{-4}$\\
& & & & & & & & & & & & \\
 max & C   & 20 & 16.63 & 0.98 & -2.57 & 0.49 & 1.86 & 0.28 & 
 0.134 & 6.2 & 0.845 & $5.21 \times 10^{-6}$\\
& & & & & & & & & & & & \\
 min & C   & 20 & 21.28 & 2.14 & -4.37 & 0.96 & 2.15 &  0.29 &
 0.337 & 15.4 & 0.874 & $9.77 \times 10^{-7}$\\
& & & & & & & & & & & & \\
mean & O   & 29 & 19.02 & 0.69 & -3.18 & 0.33 & 1.13 &  0.75 &
 0.183 & 8.4 & 0.282 & 0.145\\
& & & & & & & & & & & & \\
 max & O   & 28 & 19.46 & 0.54 & -3.51 & 0.28 & 1.11 &  0.57 &
 0.150 & 6.9 & 0.366 & $6.03 \times 10^{-2}$\\
& & & & & & & & & & & & \\
 min & O   & 28 & 19.05 & 1.11 & -2.68 & 0.48 & 0.15 &  0.74 &
 0.297 & 13.7 & 0.041 & 0.838\\
& & & & & & & & & & & & \\
mean & O+C & 49 & 18.53 & 0.63 & -3.30 & 0.31 & 2.06 & 0.19 &
 0.210 & 9.7 & 0.848 & $2.65 \times 10^{-14}$\\
& & & & & & & & & & & & \\
 max & O+C & 48 & 18.59 & 0.47 & -3.54 & 0.24 & 2.23 & 0.18 &
 0.154 & 7.1 & 0.876 & $7.05 \times 10^{-16}$\\
& & & & & & & & & & & & \\
 min & O+C & 48 & 18.84 & 0.97 & -3.21 & 0.45 & 1.84 & 0.18 &
 0.325 & 15.0 & 0.841 & $1.47 \times 10^{-13}$\\
\end{tabular}
\end{table*}

\begin{table*}
\caption[]{K band fitted PLC relations using $(J-H)$ colour}
\noindent
\begin{tabular}[t]{ccccccccccccl}
fit &  sample & $n$ & $a$ & $\sigma_a$ & $b$ & $\sigma_b$ & $c$
& $\sigma_c$ & $\sigma$ & $\Delta$ & $\hat{\rho}$ & Prob\\ \hline
& & & & & & & & & & & & \\
mean & C   & 20 & 18.21 & 1.24 & -2.84 & 0.60 & -0.29 & 0.30 &
 0.176 & 8.1 & -0.233 & 0.338\\
& & & & & & & & & & & & \\
 max & C   & 20 & 16.60 & 1.20 & -2.22 & 0.60 & -0.51 &  0.35 & 
 0.164 & 7.6 & -0.335 & 0.161\\
& & & & & & & & & & & & \\
 min & C   & 20 & 20.83 & 1.75 & -4.11 & 0.78 & 0.32 &  0.24 &
 0.273 & 12.6 & 0.307 & 0.200\\
& & & & & & & & & & & & \\
mean & O   & 29 & 19.29 & 0.47 & -3.61 & 0.22 & 0.59 &  0.51 &
 0.125 & 5.8 & 0.221 & 0.259\\
& & & & & & & & & & & & \\
 max & O   & 28 & 19.59 & 0.43 & -3.95 & 0.23 & 0.81 &  0.45 &
 0.120 & 5.5 & 0.335 & $8,69 \times 10^{-2}$\\
& & & & & & & & & & & & \\
 min & O   & 28 & 19.42 & 0.81 & -3.16 & 0.35 & -0.39 &  0.54 &
 0.217 & 10.0 & -0.142 & 0.478\\
& & & & & & & & & & & & \\
mean & O+C & 49 & 19.38 & 0.45 & -3.37 & 0.22 & -0.19 & 0.13 &
 0.148 & 6.8 & -0.200 & 0.173\\
& & & & & & & & & & & & \\
 max & O+C & 48 & 19.25 & 0.46 & -3.54 & 0.24 & 0.03 & 0.18 &
 0.150 & 6.9 & 0.028 & 0.851\\
& & & & & & & & & & & & \\
 min & O+C & 48 & 19.84 & 0.74 & -3.48 & 0.34 & -0.06 & 0.14 &
 0.249 & 11.5 & -0.069 & 0.646\\
\end{tabular}
\end{table*}

\begin{table*}
\caption[]{$m_{\rm{bol}}$ fitted PLC relations using $(J-H)$ colour}
\noindent
\begin{tabular}[t]{ccccccccccccl}
fit &  sample & $n$ & $a$ & $\sigma_a$ & $b$ & $\sigma_b$ & $c$
& $\sigma_c$ & $\sigma$ & $\Delta$ & $\hat{\rho}$ & Prob\\ \hline
& & & & & & & & & & & & \\
mean & C   & 20 & 19.42 & 0.92 & -2.26 & 0.45 & 0.27 & 0.22 &
 0.130 & 6.0 & 0.286 & 0.236\\
& & & & & & & & & & & & \\
 max & C   & 20 & 17.81 & 0.80 & -1.70 & 0.40 & 0.21 &  0.23 & 
 0.109 & 5.0 & 0.211 & 0.385\\
& & & & & & & & & & & & \\
 min & C   & 20 & 22.60 & 1.53 & -3.48 & 0.69 & 0.36 &  0.21 &
 0.239 & 11.0 & 0.389 & $9.98 \times 10^{-2}$\\
& & & & & & & & & & & & \\
mean & O   & 29 & 20.99 & 0.58 & -3.28 & 0.27 & 1.18 &  0.63 &
 0.153 & 7.0 & 0.346 & $7.14 \times 10^{-2}$\\
& & & & & & & & & & & & \\
 max & O   & 28 & 21.47 & 0.54 & -3.53 & 0.28 & 0.91 &  0.57 &
 0.150 & 6.9 & 0.307 & 0.119\\
& & & & & & & & & & & & \\
 min & O   & 28 & 21.18 & 1.13 & -2.70 & 0.49 & -0.15 &  0.76 &
 0.303 & 14.0 & -0.039 & 0.846\\
& & & & & & & & & & & & \\
mean & O+C & 49 & 20.87 & 0.44 & -3.01 & 0.22 & 0.60 & 0.13 &
 0.145 & 6.7 & 0.557 & $3.90 \times 10^{-5}$\\
& & & & & & & & & & & & \\
 max & O+C & 48 & 20.72 & 0.46 & -3.22 & 0.23 & 0.97 & 0.18 &
 0.148 & 6,8 & 0.635 & $1.65 \times 10^{-6}$\\
& & & & & & & & & & & & \\
 min & O+C & 48 & 21.28 & 0.81 & -2.90 & 0.38 & 0.28 & 0.15 &
 0.273 & 12.6 & 0.271 & $6.54 \times 10^{-2}$\\
\end{tabular}
\end{table*}

The PLC results presented here correspond to the `Method a' case presented
in F89, i.e. ordinary least squares with the errors in the magnitudes.
These solutions do not account for the effect of correlated errors on
apparent magnitude and colour excess -- an effect treated in detail by
e.g. Caldwell \& Coulson (1985). Following F89 we conclude that this effect
is negligible for these data, since the extinction in the J, H and K bands
is very small.

Our results at mean light using $(J-K)$ colours are, as expected, in complete
agreement with those of F89. Moreover our conclusions concerning the
significance of the colour term are concordant with F89, and indeed are
reinforced by considering the value of the sample multiple correlation
coefficient, $\hat{\rho}$, and associated P value. Clearly the magnitude
of the colour term itself, compared with its standard error, also gives
an indication of its statistical significance. On this basis, we can see that
there is good evidence for a $(J-K)$ PLC relation at mean, maximum and
minimum light in almost all cases considered.  The only exceptions are the
bolometric PLC relation for the C Miras (for which the P values listed
in Table 10 are at least several orders of magnitude larger than in other
cases) and the K band PLC relation for the C Miras, which -- although not
listed in the Tables -- can be seen to have a marginal colour term at mean,
maximum and minimum light by considering the $c_J$ coefficients and
using $c_K = c_J-1$ as noted above. The colour term is highly
significant for the J band relations, slightly less significant for the
H band relations and least significant (although still clearly present)
for the bolometric relations.

For the results using $(J-H)$ colours there is
clear evidence of a significant colour term for the J band `C' and `O+C'
relations and -- in view of the values of $c_J$ -- also for the H band
`C' and `O+C' relations. The colour term is also significant for the
bolometric PLC relation with the `O+C' sample. In all other cases, however,
there is no evidence for a significant colour term -- i.e. the addition
of $(J-H)$ colour does not significantly reduce the dispersion of the
PL relation.

It is clear from Tables 8 to 13 that in all cases
the dispersion of the PLC relations at minimum light is considerably larger 
than at mean and maximum light, for both $(J-K)$ and $(J-H)$ colours.
Comparing the dispersion at mean and maximum light, however, we see that our 
results are somewhat more ambiguous than was the case for the PL relations 
considered in the previous section. For $(J-K)$ colours the dispersion at
maximum light is in fact slightly {\em larger\/} than that at mean light
for the O Miras with J band, H band and bolometric magnitudes, and also for the
`O+C' sample bolometric relation. In the remaining five cases the dispersion
at maximum light is smaller than at mean light. What is noteworthy, however,
is the particular success of the J and H band maximum light relations 
for the C Miras: these are the two cases for which the dispersion at mean light
is largest, and the reduction in dispersion at maximum light is found to be
about $20 - 30 \%$.

For the PLC relations with $(J-H)$ colour, where the colour term was not
significant the ratio of the maximum to mean dispersion was very similar
to that for the corresponding PL relations -- i.e. the dispersion of the
maximum light relation was generally found to be comparable to, or slightly
smaller than, that at mean light. This is not surprising since in these
cases the fitted PLC relation shows no statistically significant difference
from the PL relation. For the cases where a statistically significant
$(J-H)$ colour term {\em was\/} found, on the other hand, the maximum light
relations scored two notable successes: the dispersion of the J band relations
for the `C' and `O+C' samples was reduced by almost 50\% and 25\% respectively.
Note that, as for the $(J-K)$ relations, the largest reduction in dispersion
at maximum light occurred for the PLC relations with largest dispersion at mean
light. Note also that for the C Miras the J band PL
dispersion at maximum light is already smaller than the corresponding
PLC relation at mean light with $(J-H)$ colour, while the maximum light 
PLC relation reduces the dispersion by almost another factor of two.

Tables 14 and 15 list the results of applying to the fitted PLC relations
the test, introduced
in the previous section, to determine the statistical significance of
the reduction (or increase!) in scatter between mean and maximum light.
The columns are as in Table 7 above. Table 14 gives the results using $(J-K)$
colours while Table 15 is for $(J-H)$ colours.

\begin{table}
\caption[]{Significance of the reduction in dispersion of maximum light PLC
relations compared with mean light PLC relations, both using
$(J-K)$ colours. Column headings are as explained in Section 3 above}
\noindent
\begin{tabular}[t]{ccccl}
      waveband & sample & $R$ & $r$ & Prob\\ \hline
               & & & & \\
             J & C  & 1.511 & 0.800 & $7.5 \times 10^{-2}$\\
             H & C  & 2.296 & 0.705 & $8.7 \times 10^{-3}$\\
$m_{\rm{bol}}$ & C  & 1.018 & 0.709 & 0.477\\
               & & & & \\
             J & O  & 0.737 & 0.597 & 0.836\\
             H & O  & 0.916 & 0.541 & 0.604\\
$m_{\rm{bol}}$ & O  & 0.854 & 0.646 & 0.696\\
               & & & & \\
             J & O+C & 1.086 & 0.667 & 0.355\\
             H & O+C & 1.430 & 0.575 & $7.2 \times 10^{-2}$\\
$m_{\rm{bol}}$ & O+C & 0.835 & 0.737 & 0.752\\
\end{tabular}
\end{table}

\begin{table}
\caption[]{Significance of the reduction in dispersion of maximum light PLC
relations compared with mean light PLC relations, both using
$(J-H)$ colours. Column headings are as explained in Section 3 above}
\noindent
\begin{tabular}[t]{ccccl}
      waveband & sample & $R$ & $r$ & Prob\\ \hline
               & & & & \\
             J & C  & 3.565 & 0.493 & $1.5 \times 10^{-3}$\\
             K & C  & 1.152 & 0.704 & 0.341\\
$m_{\rm{bol}}$ & C  & 1.422 & 0.484 & 0.192\\
               & & & & \\
             J & O  & 1.488 & 0.742 & $1.0 \times 10^{-1}$\\
             K & O  & 1.085 & 0.662 & 0.389\\
$m_{\rm{bol}}$ & O  & 1.040 & 0.811 & 0.424\\
               & & & & \\
             J & O+C & 1.860 & 0.494 & $8.4 \times 10^{-3}$\\
             K & O+C & 0.974 & 0.672 & 0.535\\
$m_{\rm{bol}}$ & O+C & 0.960 & 0.679 & 0.561\\
\end{tabular}
\end{table}

The results of Tables 14 and 15 confirm that in the cases where the
dispersion at maximum light is greater than at mean light
($\Delta < 1$) the increase in dispersion is never significant at less than
the 15\% level, while the {\em reduction\/} in dispersion at maximum light
is in several other cases significant at the 1\% level.

\section{Discussion}

The principal result of this paper concerns the existence of Mira PL
relations at maximum light and the fact that in all cases these relations
display less scatter than the corresponding mean light relations, a
reduction in dispersion which is statistically significant in the J and H
bands. This result
is apparent not only in the values of the dispersion derived for the
mean and maximum light relations, but also is suggested by the
behaviour of some of the outliers in the scatterplots of
magnitude against log period. In Figures 3 and 4 for example, there are
a small number of outliers, with log periods of around 2.5,
in the J and H band mean light relations which are in much better agreement
with the fitted regression line at maximum light. We also find that in
several cases for the C Miras the dispersion of the maximum light PLC
relation is significantly smaller, by up to 50\%, than at mean light.

It is obviously important now to ask what is the most likely source of the
reduction in dispersion which we have observed
at maximum light. Figures 7 and 8 show plots of
dereddened $(J-K)$ colour against log period at mean (a) and maximum (b)
light, for the C and O Miras respectively.
(Note that `maximum $(J-K)_0$', as indicated on the axes of the plots in
Figures 7 and 8, in fact means the dereddened colour at the phase of
maximum light in the J band, which need not be the same as the maximum
observed value of J-K colour, although the difference is likely to be quite
small). Similarly Figures 9 and 10 show plots of dereddened
$(J-H)$ colour against log period for the C and O Miras respectively.

\begin{figure}
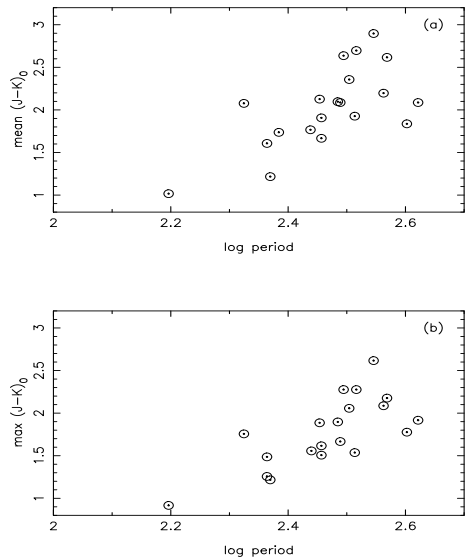

\centerline{\psfig{file=jmk_cmean.eps,width=7cm,height=4cm,angle=-90}}
\centerline{\psfig{file=jmk_cmax.eps,width=7cm,height=4cm,angle=-90}}
\caption[]{Scatterplots of log period against dereddened $(J-K)$ colour
at (a) mean and (b) maximum light, for the sample of carbon-rich Miras}
\label{fig:fig7}
\end{figure}

\begin{figure}
\centerline{\psfig{file=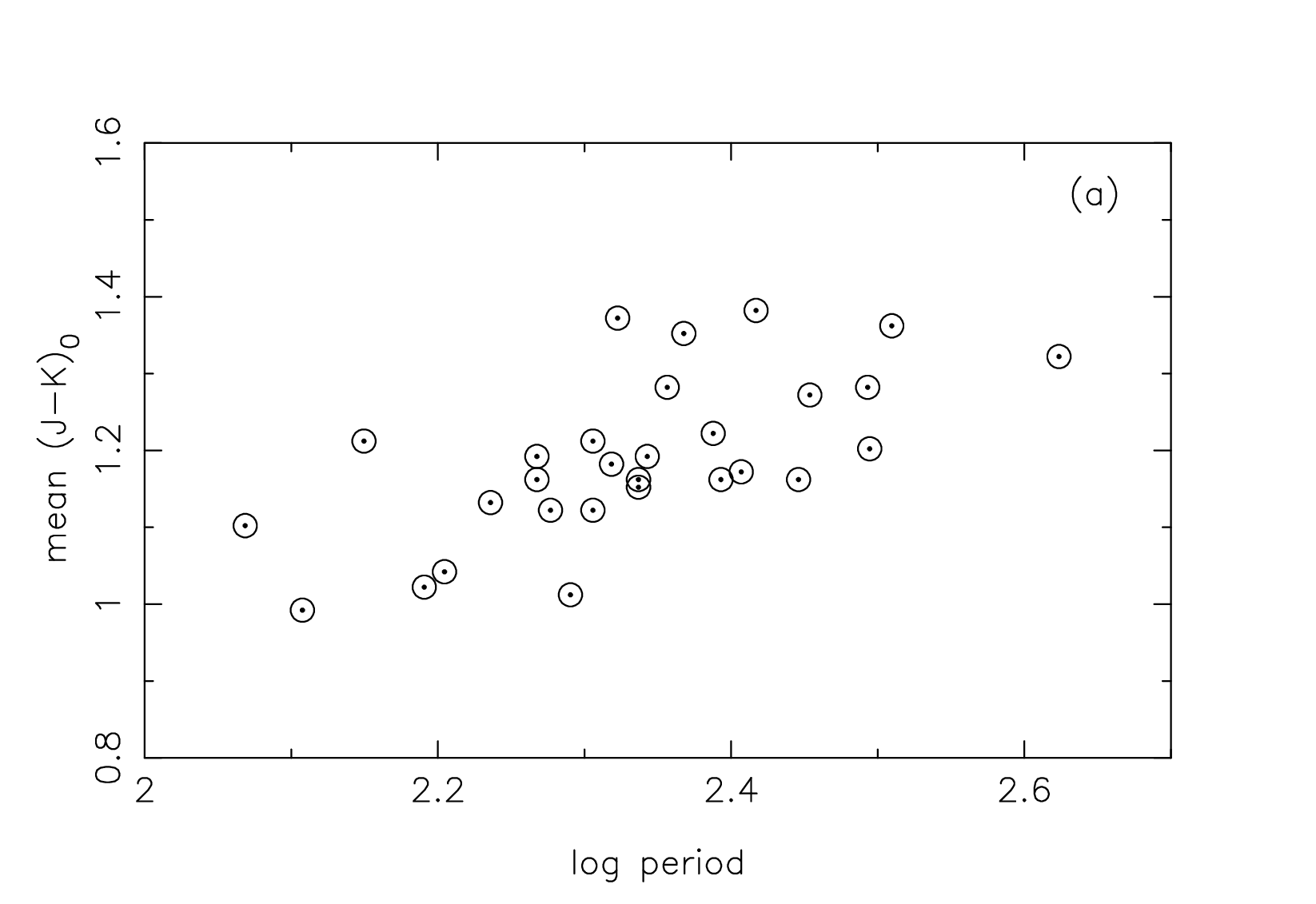,width=7cm,height=4cm,angle=-90}}
\centerline{\psfig{file=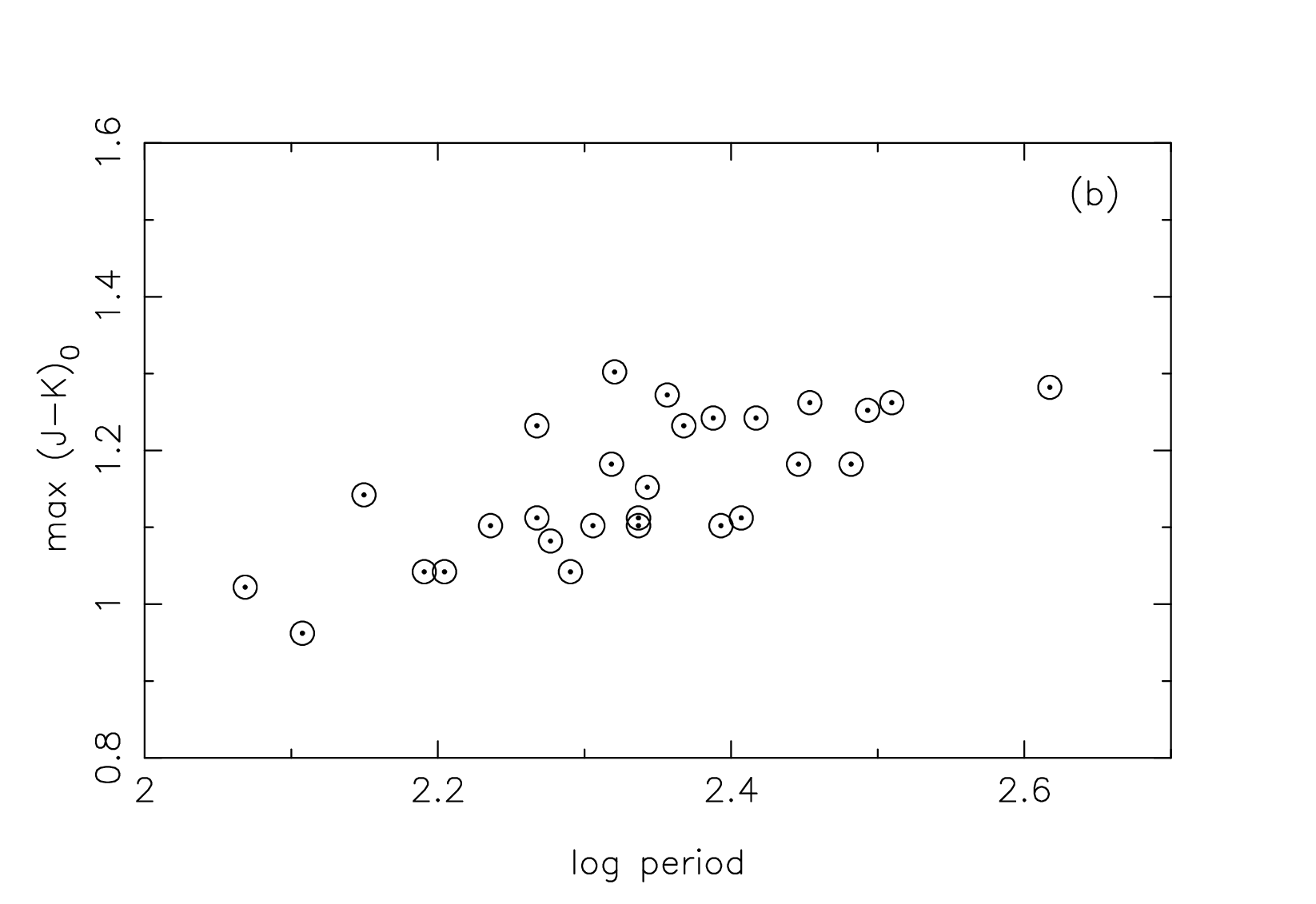,width=7cm,height=4cm,angle=-90}}
\caption[]{Scatterplots of log period against dereddened $(J-K)$ colour
at (a) mean and (b) maximum light, for the sample of oxygen-rich Miras}
\label{fig:fig8}
\end{figure}

\begin{figure}
\centerline{\psfig{file=jmh_cmean.eps,width=7cm,height=4cm,angle=-90}}
\centerline{\psfig{file=jmh_cmax.eps,width=7cm,height=4cm,angle=-90}}
\caption[]{Scatterplots of log period against dereddened $(J-H)$ colour
at (a) mean and (b) maximum light, for the sample of carbon-rich Miras}
\label{fig:fig9}
\end{figure}

\begin{figure}
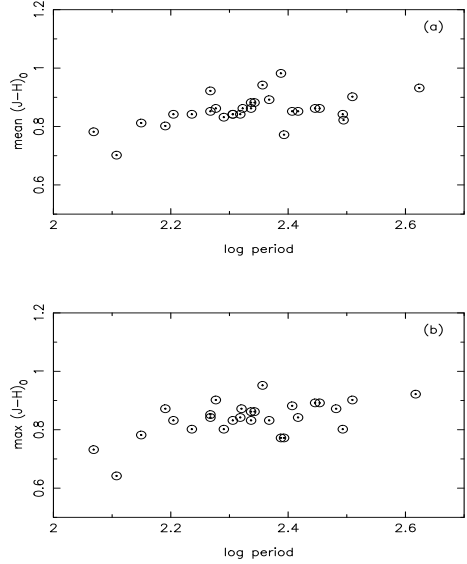

\centerline{\psfig{file=jmh_omean.eps,width=7cm,height=4cm,angle=-90}}
\centerline{\psfig{file=jmh_omax.eps,width=7cm,height=4cm,angle=-90}}
\caption[]{Scatterplots of log period against dereddened $(J-H)$ colour
at (a) mean and (b) maximum light, for the sample of oxygen-rich Miras}
\label{fig:fig10}
\end{figure}

Whilst the properties of Mira PLC relations at maximum light will be 
the focus of future work, we note from these plots that the scatter in
$(J-K)_0$ at a given period is very similar at mean and maximum light. This
suggests that the range of effective temperatures at given period will
not be greatly different at mean and maximum light, as was also claimed by
Feast (1995). In the light of our discussion in Section 1, we therefore
conjecture that the smaller dispersion of maximum light PL relations
compared with mean PL relations is primarily due to the fact that both
maximum luminosity and period depend on the total mass, 
whereas equilibrium luminosity -- and hence mean luminosity -- 
depends strongly on core mass. If the above explanation is correct then --
insofar as a considerably larger reduction in
dispersion is found in the J and H bands for the C Miras compared with the
O Miras -- one might suppose this to be due to there being a larger difference
between the range of core masses and total masses for C Miras compared with
that for O Miras. It would be interesting to investigate this possibility
further, and carry out a more detailed study of the systematic differences
between C and O Miras.

In comparing the values of the regression coefficients obtained in the 
PL fits at mean and maximum light, the general trend which one observes is
as follows. For the C Miras the zero point, $b$, and slope, $a$, are
found to be smaller and more shallow (i.e. less negative) respectively
for the maximum light relations in all wavebands. Based on the standard
errors of the regression
coefficients this systematic difference appears to be quite significant,
although we have not carried out a specific statistical test of this
hypothesis. For the O Miras, on the other hand, precisely the converse 
is the case: the zero point and slope of the PL relations at maximum light
are found to be respectively larger and steeper (i.e. more negative)
than at mean light in all wavebands. Aside from investigating whether the
use of the maximum light relations derived in this study lead to distance
estimates significantly different from those determined using Mira PL
relations at mean light (c.f. F89, Whitelock 1995, Feast 1995), it
would be interesting to investigate if any systematic difference in the
slope and zero point of maximum light relations can be explained in terms
of our existing knowledge of the physics of Mira variables. We will
address this problem further in a future paper.

In the case of Cepheids it was shown in Simon, Kanbur \& Mihalas (1993) that
{\em both\/} maximum and minimum light occur as the star is passing
through its equilibrium radius. If this were also true for Miras, then
equation (3) might also be valid at minimum light, but with
$L_{\rm{max}}$ and $T_{\rm{max}}$ replaced by $L_{\rm{min}}$ and
$T_{\rm{min}}$. As we commented above, the Mira PL and PLC relations at
minimum light were in all cases found to have a larger dispersion than at
mean light. In view of our discussion
in Section 1, perhaps one reason for this result is that minimum
luminosity is dependent on the total mass of the star in a different way to
maximum luminosity; in other words it may be the case that equation (3) --
in its equivalent form -- does indeed hold at minimum light but that when we 
collapse the equation over the variables $\log {\cal{M}}$ and
$\log T_{\rm{min}}$ the resultant PL relation has a larger dispersion than at
either maximum or mean light. We intend to investigate further the properties
of PL relations at minimum light in future work.

Whilst the mean and maximum light PL relations which we have derived here
clearly have practical use in terms of in terms of distance estimation,
the relations involving bolometric magnitudes are also very
interesting for the purpose of better understanding stellar pulsation and 
evolution. Indeed, the fact that we have established bolometric
PL relations for both C and O Miras directly supports the validity of
equation (3). 

Finally it is important to comment explicitly
on the practical application of the results of this paper --
the use of maximum light PL relations for Miras as distance
indicators. Aside from the advantage of the small reduction in
dispersion which our analysis in this paper has identified, the use of
maximum light relations in distance estimation can also be justified on
the grounds that one can extend their application to greater distances
before the effects of luminosity selection bias become important. In
Hendry, Kanbur \& Clarke (1997, in prep.) we examine in detail the
robustness of PL relations derived for Miras detected close
to an apparent magnitude limit, and find that -- as one approaches the
magnitude limit -- the measurement of mean light becomes biased, and is
subject
to a increasingly large root mean squared error, substantially more quickly
than does maximum light. Moreover, for a range of different light
curve shapes we find that the identification of maximum light simply
with the brightest observed phase point (as was the definition adopted
in this paper) remains a robust and reliable estimate of maximum light
as one approaches the magnitude limit -- provided one has of the order of
ten or more sampled phase points -- and is certainly considerably more
robust than the identification of mean light with the average of the
observed magnitudes. This work suggests that maximum light PL relations
can easily be constructed without recourse to exhaustive observing
programmes and can therefore prove useful in extending the
range and reliability of Mira-based distance indicators. It would seem
to us, therefore, that a
priority for future work is to establish the existence of Mira PL
relations at maximum light in different stellar environments, such as the
SMC and Galactic globular clusters, and to test the uniformity of such
relations.

Wood (1990) has used equations for the position of AGB in the  
HR diagram together with the period-mean density theorem
to obtain a pulsation equation similar to equation (1), but also
incorporating a metallicity dependence. This work is discussed in Feast 
(1995), which suggests that the available evidence indicates little
variation in the mean light Mira PL relations at K or bolometric magnitudes 
in environments with a range of different metallicities. There is 
no reason to believe why any metallicity dependence of equation (1)
would act differentially between mean and maximum light, although of course
any possible effect should certainly be checked observationally. In any case,
such a metallicity gradient with environment -- if present -- would have
no bearing on our discussion of the relative dispersion of mean and maximum
light PL relations in this paper.

Some examples of the recent application of distance indicators based on
Mira PL relations at mean light include the following. In F89 Mira distances
were determined to Galactic globular clusters, thus providing an absolute
calibration of RR Lyrae and horizontal branch stars. On the other hand,
Whitelock (1995) and references therein used Mira PL relations at mean light
to study the dimensions and kinematics of the disk, halo and bulge of the
Galaxy.

\section{Conclusions}

In this paper we have demonstrated the existence of
PL relations for Miras at maximum light in the
J, H and iK bands and for bolometric magnitudes. Our results were based
on analysis of a sample of oxygen-rich and carbon-rich Miras in the LMC,
as previously studied in F89. In the J, H and K bands the PL relations at
maximum light have a smaller dispersion for the oxygen-rich
Miras than for the carbon rich Miras, while the converse was found to be true for the PL relation
based on bolometric magnitudes. We have shown that for the J and H bands
the Mira PL relations at maximum light have a significantly smaller
dispersion than their 
counterparts at mean light. Our results also are suggestive that C Miras with 
periods in excess of 250 days have constant mean and maximum absolute
magnitude.

Based on similar reasoning to that outlined in KH, we present a
theoretical justification for the existence of such
maximum light relations. The crucial assumption made in this justification
is that the photospheric radius at mean light is roughly equal to
the photospheric radius at maximum light, for which there exists some
evidence. Amongst other factors -- including metallicity and temperature --
the dispersion at given period in a mean light PL relation is influenced by 
both the range of core masses and the range of total masses found in Miras.
However, at maximum light we suggest that -- amongst these same other
factors -- the dispersion at given period is influenced only by the range of 
total masses, and that it is this fact which is responsible for
the smaller dispersion of maximum light PL relations which we have observed.

In Section 5 we have outlined a number of topics for future work, but
in summary it seems clear that the main direction of future work should
be the study of larger samples of Miras, in order to investigate the
prevalence, uniformity and reliability of maximum light relations in
other environments. The relative robustness of maximum light relations
when the corresponding mean light relations are pushed close to an
apparent magnitude limit makes their application in external galaxies
an important and exciting possibility -- particularly with the installation
of J and H band filters in the NICMOS camera on the newly refurbished
Hubble Space Telescope.
We are confident that Mira PL relations at maximum light can
become a powerful tool for galactic and extragalactic astronomy.
\vspace{5mm}

\noindent
{\sc{acknowledgements}}
The authors thank Patricia Whitelock for supplying the Mira observations
of F89 in a convenient electronic form, and Shaun Hughes, Tom Lloyd
Evans, Norman Simon and Dimitri Mihalas for useful discussions. The authors
also thank the anonymous referee for useful comments.
MAH acknowledges the PPARC, for the award of a Personal Research
Fellowship. MAH and SMK ackowledge the use of computer facilities supported
by the STARLINK project. DC acknowledges the assistance of Mrs. Margaret
Morris in analysing the data.

\end{document}